\documentclass{aa}
\usepackage{booktabs}
\usepackage{amsmath}
\usepackage{txfonts}
\usepackage{graphicx}
\usepackage{txfonts}
\newcommand{\cii}{[C\,{\sc ii}]}
\newcommand{\sii}{[S\,{\sc ii}]}
\newcommand{\feii}{[Fe\,{\sc ii}]}
\newcommand{\hii}{\ion{H}{ii}}
\newcommand{\oi}{[O\,{\sc i}]}
\usepackage{hyperref}
\hypersetup{
  colorlinks = true, %Colours links instead of ugly boxes
  urlcolor   = blue, %Colour for external hyperlinks
  linkcolor  = blue, %Colour of internal links
  citecolor  = blue %Colour of citations
}

\begin{document}

   \title{Breaking Orion's Veil bubble with fossil outflows}
   %Punching the Orion's Veil bubble with fossil outflows
   %Breaking Orion's Veil bubble via outflows of Trapezium cluster
   %\subtitle{I. Overviewing the $\kappa$-mechanism}

   \author{\"U.~Kavak\inst{1,2,3,4}
          \and J.~R.~Goicoechea\inst{5}
          \and C.~H.~M.~Pabst\inst{4,5}
          \and J.~Bally\inst{6}
          \and F.~F.~S.~van~der~Tak\inst{2,1} 
          \and A.~G.~G.~M.~Tielens\inst{4}
          }

   \institute{Kapteyn Astronomical Institute, University of Groningen, P.O. Box 800, 9700 AV Groningen, The Netherlands \\
             \email{ukavak@sofia.usra.edu}
         \and
             SRON Netherlands Institute for Space Research, Landleven 12, 9747 AD Groningen, The Netherlands
         \and
             SOFIA Science Center, USRA, NASA Ames Research Center, M.S. N232-12, Moffett Field, CA 94035, USA
         \and
              Leiden Observatory, Leiden University, PO Box 9513, NL-2300RA, Leiden, the Netherlands
         \and
             Instituto de F\'isica Fundamental, CSIC, Calle Serrano 121-123, 28006 Madrid, Spain
         \and
            Department of Astrophysical and Planetary Sciences, University of Colorado, Boulder, Colorado 80389, USA
             }

   \date{Received Month XX, 2021; accepted Month XX, 2021}

% \abstract{}{}{}{}{} 
% 5 {} token are mandatory
 
  \abstract
   {The role of feedback in the self-regulation of star formation is a fundamental question in astrophysics. The Orion Nebula is the nearest site of ongoing and recent massive star formation. It is a unique laboratory for the study of stellar feedback. Recent SOFIA \cii\,158~$\mu$m observations revealed an expanding bubble, the Veil shell, being powered by stellar winds and ionization feedback.} 
   % aims heading ( mandatory )
   {We have identified a protrusion-like substructure in the Northwest portion of the Orion Veil Shell that may indicate additional feedback mechanisms that are highly directional. Our goal is to investigate the origin of the protrusion by quantifying its possible driving mechanisms.}
   {We use the \cii\,158~$\mu$m map of the Orion Nebula obtained with the upGREAT instrument onboard SOFIA. The spectral and spatial resolution of the observations are 0.3~km~s$^{-1}$ and $16\arcsec$, respectively. The velocity-resolved \cii\,observations allow us to construct position-velocity (pv) diagrams to measure the morphology and the expansion velocity of the protrusion. For the morphology, we also use new observations of $^{12}$CO and $^{13}$CO $J$ = 2-1 (to trace molecular gas), \textit{Spitzer} 8~$\mu$m (to trace the far-UV illuminated surfaces of photodissociation regions), and H$\alpha$ (to trace ionized gas). For the kinematics, we perform line-profile analysis of \cii, $^{13}$CO, and $^{12}$CO at twelve positions covering the entire protrusion. To quantify the stellar feedback, we estimate the mass of the protrusion by fitting the dust thermal emission. We compare the kinetic energy with the stellar wind of $\theta^1$~Ori~C and the momentum of the outflows of massive protostars to investigate the driving mechanism of the protrusion.} 
   % results heading ( mandatory ) 
   {The pv diagrams reveal two half-shells expanding at velocities of $+$6~km~s$^{-1}$ and $+$12~km~s$^{-1}$. We find that the protrusion has a diameter of $\sim$1.3~pc with a $\sim$45~$M_\odot$ shell expanding at $+$12~km~s$^{-1}$ at the northwestern rim of the Veil. The thickness of the expanding shell is $\sim$0.1~pc. Using the mass in the limb-brightened shell and the maximum expansion velocity, we calculate the kinetic energy and the momentum of the protrusion as $\sim$7~$\times$~10$^{46}$~erg and 540~$M_\odot$~km~s$^{-1}$, respectively. We consider three possible origins for this protrusion: Fossil outflow cavities created by jets/outflows during the protostellar accretion phase, pre-existing clumpiness in the OMC-1 core, and the stellar wind during the main sequence phase. Based on the energetics and the morphology, we conclude that the northwestern part of the pre-existing cloud was locally perturbed by outflows ejected from massive protostars in the Trapezium cluster. This suggests that the protrusion of the Veil is the result of mechanical rather than radiative feedback. Furthermore, we argue that the location of the protrusion is a suitable place to break the Orion Veil owing to the photo-ablation from the walls of the protrusion.
   }
   {We conclude that the outflows of massive protostars can influence the morphology of the future \hii\,region and even cause breakages in the ionization front. Specifically, the interaction of stellar winds of main-sequence stars with the molecular core pre-processed by the protostellar jet is important.
   }
   \keywords{Stars: massive -- 
             ISM: bubbles -- 
             ISM: kinematics and dynamics}

   \maketitle
%

%-------------------------------------------------------------

\section{Introduction}

    Massive stars have luminosities larger than 10$^3$~L$_\odot$, corresponding to a spectral type of B3 or earlier, and have stellar masses higher than 8~$M_\odot$. The formation of massive stars is far less understood than that of low-mass stars \citep[< 8~$M_\odot$; see reviews by][]{Tan2014, Motte2018}. Forming massive stars differs from forming low-mass stars in several ways. Their Kelvin-Helmholtz times are much shorter owing to their high luminosities. They tend to form in dense clusters and exhibit a higher multiplicity fraction \citep{Motte2018}. While accreting at high rates, massive stars growing through 10 to 15~$M_\odot$ develop extended photo-spheres resembling red giants \citep{Hosokawa2009}. Recent studies examine the formation of massive stars and its similarity to low-mass star formation by searching ubiquitous phenomena found in low-mass star-forming regions (such as disks, jets, and outflows in the scenario of disk-mediated accretion; see \citealt{Beuther2002, LopezSepulcre2010, SanchezMonge2013, Cesaroni2017, Purser2018, Sanna2018, Kavak2021}). Massive stars, in contrast to low-mass stars, reach their main-sequence luminosity while still embedded in accreting a natal cloud of gas and dust \citep{Hosokawa2009, Kuiper2011}. A massive protostellar embryo heats and ionizes the gas of its surrounding envelope with Extreme Ultraviolet photons (EUV; E>13.6 eV), creating an \hii\,region \citep{Spitzer1978}. Young massive stars are surrounded by ultracompact (UC) \hii\,regions with size <~0.1~pc and density >~10$^4$~cm$^{-3}$ \citep{Churchwell2002}.
    
    The gas in the UC\hii\,region is photoionized and heated by EUV photons leading to an increase in gas pressure. This highly pressurized gas causes the \hii\,region to expand until it reaches an equilibrium Str\"omgen sphere with a much lower gas density \citep{Newman1968}. In the standard model of \hii\,region evolution \citep{Spitzer1978}, the thermal pressure of the plasma drives a D-type shock into the surrounding neutral medium that sweeps-up a dense, expanding shell which traps the ionization front or photodissociation region \citep[or PDR; see review by][]{Tielens1985, Hollenbach1997, Wolfire2003}. \hii\,regions are mainly classified on the basis of their size and internal density \citep{Kurtz2005}, which span orders of magnitude in size (from 0.02 to 100~pc) and electron density (from 10 to 10$^6$~cm$^{-3}$). In addition, \hii\,regions are associated with interstellar bubbles due to their spherical morphology. The mid-IR Galactic Legacy Infrared Mid-Plane Survey Extraordinaire (GLIMPSE), obtained with NASA's \textit{Spitzer} Space Telescope, revealed parsec-sized bubbles throughout the Galactic plane \citep{Churchwell2006}\footnote{The GLIMPSE survey has revealed the omni presence of bubbles in the ISM \citep{Churchwell2006}. Most of these reflect the presence of \hii\,regions \citep{Anderson2014}. The \hii\,region bubble morphology may be driven by the thermal expansion of gas ionized by a centrals star, by the activity of a stellar wind during the main sequence phase or fossil cavities created by now extinct energy and momentum sources such as protostellar outflows. Throughout the paper, we presume that the Orion Nebula is mainly blown-up by stellar winds from the Trapezium stars \citep{Pabst2019}.}. %\citet{Krumholz2009} showed that bubble expansion driven only by ionized gas is insufficient and that other mechanisms than the pressure of the photoionized gas are needed to reproduce giant molecular clouds (GMCs).
 
    %First describe the importance of the stellar feedback and what it is?
    Stellar feedback results from the injection of energy, momentum, and mass into the interstellar medium (ISM) by massive stars. This feedback is a combination of ionizing radiation, radiation pressure, stellar winds, and supernovae on various spatial scales (from $\sim$1 to $\sim$100~pc) and dynamical timescales (from 10$^4$ to 10$^6$~years). Without stellar feedback, the temperature of interstellar matter drops rapidly, and as a consequence of this cooling, new stars form rapidly by consuming the available gas content in the Galaxy \citep{Keres2009, NaabOstriker2017, Lopez2014}. By heating up the gas and injecting turbulence in star-forming regions, stellar feedback plays a key role in preventing this `cooling catastrophe' in the evolution of galaxies \citep{Ceverino2009, Walch2012, Genzel2015}. 
    
    Feedback processes are divided into momentum- and energy-driven mechanisms which have different efficiencies in terms of energy input and time ranges \citep{Fierlinger2016}. For example, feedback from supernovae could provide enormous energy input that can shape the content of galaxies on large scales (10$-$100~pc), but much of that energy may be expended in rejuvenating hot gas in supernova remnants rather than coupling to molecular gas. On the other hand, pre-SN feedback is also crucial to reproduce the properties of GMCs \citep{Fujimoto2019, Olivier2020}. In the last two decades, observational studies have also demonstrated that feedback mechanisms have an important role in the dynamics of star forming regions \citep{Lopez2011, NaabOstriker2017}.
    
    Wind bubbles produced by stars of spectral-type earlier than B2 are described by \citet{Castor1975} and subsequently studied analytically by \citet{Weaver1977}. However, the expansion of the bubbles, in other words, their main driving feedback mechanism and the underlying physical process, are poorly understood, but are studied by simulations, which are capable of incorporating several types of feedback mechanisms individually \citep{Walch2012, Haid2018}. In the past, it has been notoriously difficult to assess the relative contribution of expansion, observationally, but that is now rapidly changing with large scale \cii\,surveys enabled by SOFIA\footnote{Stratospheric Observatory for Infrared Astronomy or SOFIA is a Boeing 747SP aircraft modified to carry a 2.7-meter telescope \citep{Young2012}.}.
    
    Most commonly, the neutral gas in the shells that confine these bubbles is translucent to far-UV (FUV) dissociating radiation, thus they host little CO to be detected \citep[e.g.,][]{Goicoechea2020} because CO is readily dissociated at low $A_\mathrm{V}$. In addition, most stars lie in the atomic or ionized phases of the ISM and not in molecular clouds. Thus their feedback mostly impacts atomic or ionized gas not traced by molecules such as CO, as well as CO-dark H$_2$ gas \citep{Grenier2005}. To date, a few alternative tracers have been reported to probe the CO-dark H$_2$ gas (e.g., CF$^+$ $J$ = 1-0 by \citet{Guzman2012}, HF $J$ = 1-0 by \citet{Kavak2019}). However, both species produce faint emission lines and require long integration times in the various regimes of the ISM. In addition to these molecular tracers, \cii\,has been proposed as a more suitable tracer because its fine-structure transition ($^2$P$_{3/2}$\,$\to$\,$^2$P$_{1/2}$ at 158~$\mu$m or 1.9~THz, i.e., $\Delta$E/$k_B$ = 91.2~K) is the main cooling agent of predominantly neutral interstellar gas \citep{Bennett1994, Hollenbach1997}. As this \cii\,line directly probes gas exposed to the FUV photons from massive stars, the \cii\,158~$\mu$m line is an ideal tracer of many types of stellar feedback mechanisms. The \cii\,line is also one of the brightest lines in PDRs and 30\% of total \cii\,emission in the Galaxy comes from dense FUV-illuminated gas \citep{Bennett1994, Pineda2014}. Moreover, velocity-resolved observations of the \cii\,line are an excellent probe of the kinematic and physical conditions of extended PDR gas \citep{Goicoechea2015, Pabst2019}, in our case, bubble shells. Unfortunately, its rest-frame emission is not accessible from ground-based observatories. With the upGREAT instrument on-board SOFIA, it is possible to observe this transition from the stratosphere \citep{Risacher2018}. Therefore, \cii\,observation of regions with a range of massive star formation activity with stars of different spectral types will provide invaluable input for simulation of the Galaxy evolution \citep[see SOFIA/FEEDBACK Survey\footnote{FEEDBACK is a SOFIA (Stratospheric Observatory for Infrared Astronomy) legacy program dedicated to study the interaction of massive stars with their environment. It performs a survey of 11 galactic high mass star forming regions in the 158~$\mu$m (1.9~THz) line of \cii\,and the 63~$\mu$m (4.7~THz) line of \oi.};][]{Schneider2020}.
    
    \begin{figure*}[!ht]
        \centering
        \begin{tabular}{cc}
        \begin{minipage}{0.49\textwidth} \includegraphics[width=0.99\textwidth]{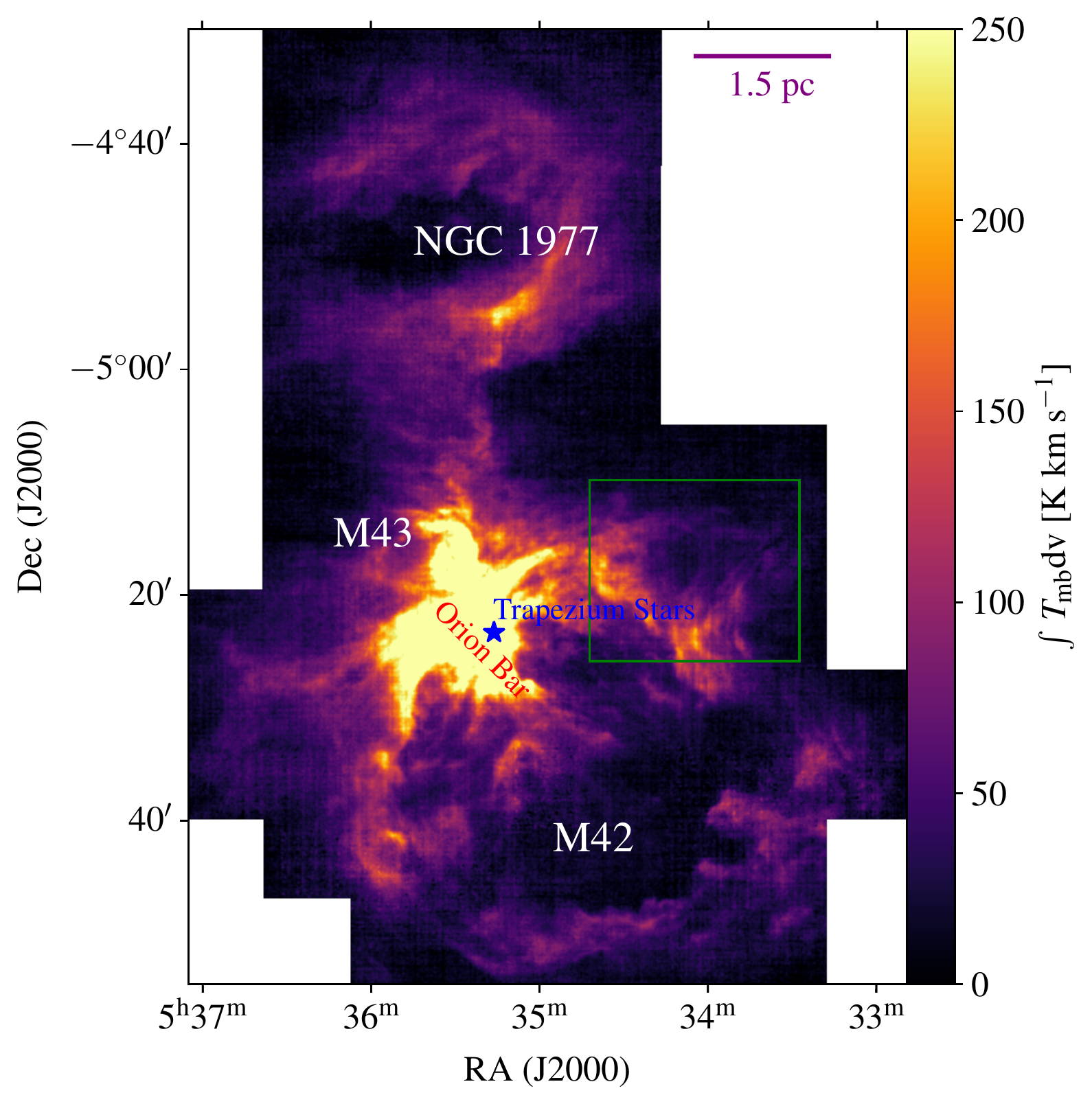} 
        \end{minipage}  
        \begin{minipage}{0.49\textwidth} \includegraphics[width=0.99\textwidth]{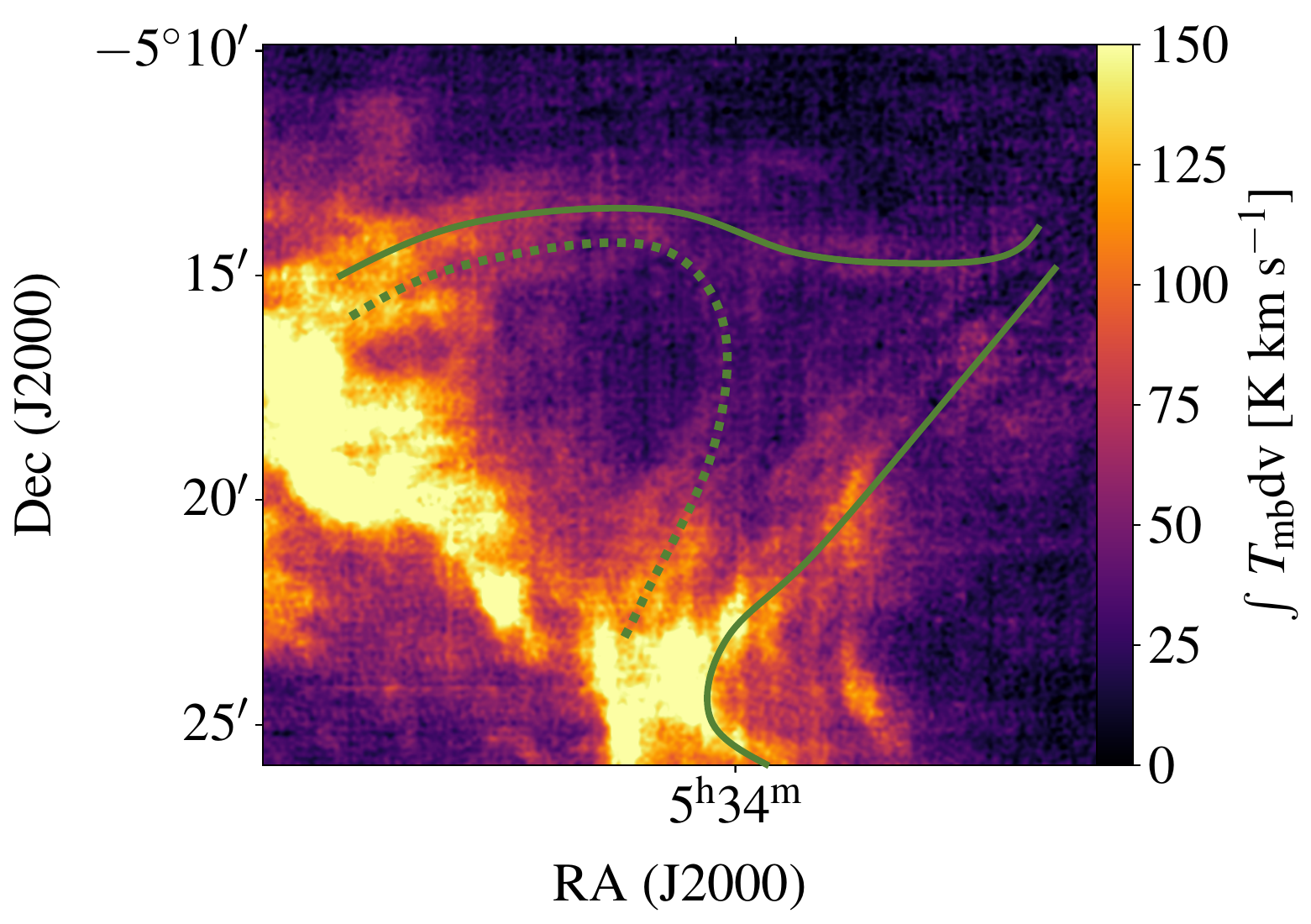}  
        \end{minipage}
    \end{tabular}
        \caption{\textit{Left:} The integrated (between $-5$ and $+$14~km s$^{-1}$) intensity \cii\,158 $\mu$m map of Orion Molecular Cloud observed by upGREAT receiver on board SOFIA. The positions of NGC~1977, Trapezium stars, M42, M43, and the Orion Bar PDR are labelled. The green box shows the extracted region from the map including the area of interest for this study, the protrusion. \textit{Right:} Close-up view of the protrusion. The bright ridge of emission is the edge of the expanding Veil shell. Faint emission extends well beyond this shell – the protrusion which has a multi component structure. See Section~\ref{Sect:Results} for more detail about the kinematics and components of the protrusion.}
    \label{fig:protsuion_cii}
    \end{figure*}
    
    Orion's Veil (Veil for short) is a series of foreground layers of gas and dust lying in front of the Trapezium stars along the line of sight towards the Orion Nebula \citep{ODell2018, Abel2019}. The Veil is a unique laboratory to study the relative effects of feedback mechanisms, as its proximity allows us to resolve the bubbles in the Orion Molecular Cloud (OMC) spatially and spectrally. Recent SOFIA \cii\,158~$\mu$m observations of the Veil focusing on the large scale emission and dynamics have shown that stellar winds have swept up the surrounding material and created the Veil shell, a half-shell of neutral gas and a mass of $\sim$1500~$M_\odot$ that expands at $\sim$13~km~s$^{-1}$ \citep{Pabst2019, Pabst2020}. They also find that stellar winds are more effective in disrupting OMC$-$1 than photo-ionization, evaporation, or even a future supernova explosion. The stellar wind is shocked, creating a hot and very dilute plasma observed in X$-$rays with Chandra \citep{Gudel2008}. The high pressure of this hot plasma has driven a shock into the environment that has swept up a dense, expanding shell of gas. In this paper, we zoom into a specific expanding structure at the north-west of the Veil using \cii\,observations. This protrusion is clearly seen in \textit{Herschel} PACS (70 and 160~$\mu$m) and SPIRE (250, 350, and 500~$\mu$m), and in \textit{Spitzer} 8~$\mu$m emission images. In this study, we investigate the origin of the protrusion using velocity-resolved SOFIA \cii\,maps and compare them to the dust, CO and PAH emission. Finally, we use the energetics of the protrusion to assess the driving mechanism.
    
    We organize the paper as follows. In Section~\ref{Section:Observation} we describe the observations of \cii, $^{12}$CO, and $^{13}$CO as well as dust emission. In Section~\ref{Sect:Results} we derive observational results on the general morphology, emission features, stars (YSO and early O$-$, B$-$, and A$-$stars) in the Veil. Section~\ref{Section:Analysis} contains a detailed analysis of the morphology, the expanding shell and its velocity, and calculations of the kinetic energy of the protrusion. Finally, we discuss whether or not the Veil is breached at the location of the protrusion in Section~\ref{Section:BreakingVeil}.

%--------------------------------------------------------------------
\section{Observations}\label{Section:Observation}
        
    \subsection{\cii\,Observations}

    The observations were conducted with the Stratospheric Observatory for Infrared Astronomy (SOFIA), which is an airborne observatory project of the US National Aeronautics and Space Administration (NASA), and the German Aerospace Centre (DLR). SOFIA is a modified aeroplane of the type Boeing 747-SP, which carries a telescope with a diameter of 2.7~m in the rear fuselage \citep{Young2012}. By flying up to 45000~ft, SOFIA makes it possible to observe at frequencies blocked by the atmosphere from the ground. A large part of the spectrum at far infrared (FIR) frequencies (1-10~THz) becomes accessible. At the same time, a few molecular species (H$_2$O, O$_3$) in the Earth's atmosphere still block FIR radiation at certain frequencies \citep{Risacher2016}.

    The data were collected with the German REceiver for Astronomy at Terahertz Frequencies (upGREAT) Instrument onboard SOFIA \citep{Risacher2018} for the Large program of the C$^+$ SQUAD led by A.~G.~G.~M.~Tielens. GREAT is a heterodyne array receiver with 21~pixels. At the time of the observations it was 2~$\times$~7 LFA plus 1~$\times$~7 pixel HFA. 2~$\times$~7-pixel sub-arrays with a hexagonal layout are designed for the low-frequency array receiver (LFA) with dual-band polarization. These cover the 1.83-2.07~THz frequency range where the \cii\,158~$\mu$m and [O\,{\sc i}] 145~$\mu$m lines can be found. The other hexagonal 7-pixel array is located in the high-frequency array (HFA) that covers the [O\,{\sc i}] 63~$\mu$m line. The GREAT instrument uses local oscillators (LO) to achieve very high spectral resolution ($\nu$/$\Delta \nu$ = 10$^7$). An area of about 1 square degree in Orion was surveyed in the \cii\,1.9 THz line \citep[cf. Fig.~\ref{fig:protsuion_cii};][]{Pabst2019}. The native spectral resolution of the map is about 0.04~km~s$^{-1}$. The final data is resampled to 0.3~km~s$^{-1}$ to achieve a better signal-to-noise ratio. The final rms noise (in $T_\mathrm{mb}$) is 1.14~K in 0.3~km~s$^{-1}$ velocity channels. The spatial resolution of the map is $16\arcsec$, which corresponds to 0.03~pc at the distance of Orion, 414 pc\footnote{We use 414~pc provided by \citet{Menten2007} as the distance. The Orion Molecular cloud does show a substantial distance gradient \citep{Grossschedel2018} but that is on a much larger scale and not relevant for our paper.} \citep{Menten2007}. The data cube is made at LSR velocities between $-$50 and $+$50~km~s$^{-1}$. The \cii\,emission mostly appears between $-$10 and $+$15~km~s$^{-1}$ in the entire cube. More detailed information about the observations has been given in \citet{Pabst2019}.
    
    We extract the \cii\,observations within the green box from the map presented in Fig.~\ref{fig:protsuion_cii}. The map is centered on an arbitrary point, that is, $\alpha$ = 05$^\mathrm{h}$34$^\mathrm{m}$17.77$^\mathrm{s}$, $\delta$ = -05$^{\degr}$20$^{\arcmin}$ 03.89$\arcsec$ (J2000) and covers the entire protrusion at the north-east of the Veil (Fig.~\ref{fig:protsuion_cii}).
       
    \begin{figure*}[!h]
    \centering
        \includegraphics[width=0.95\textwidth]{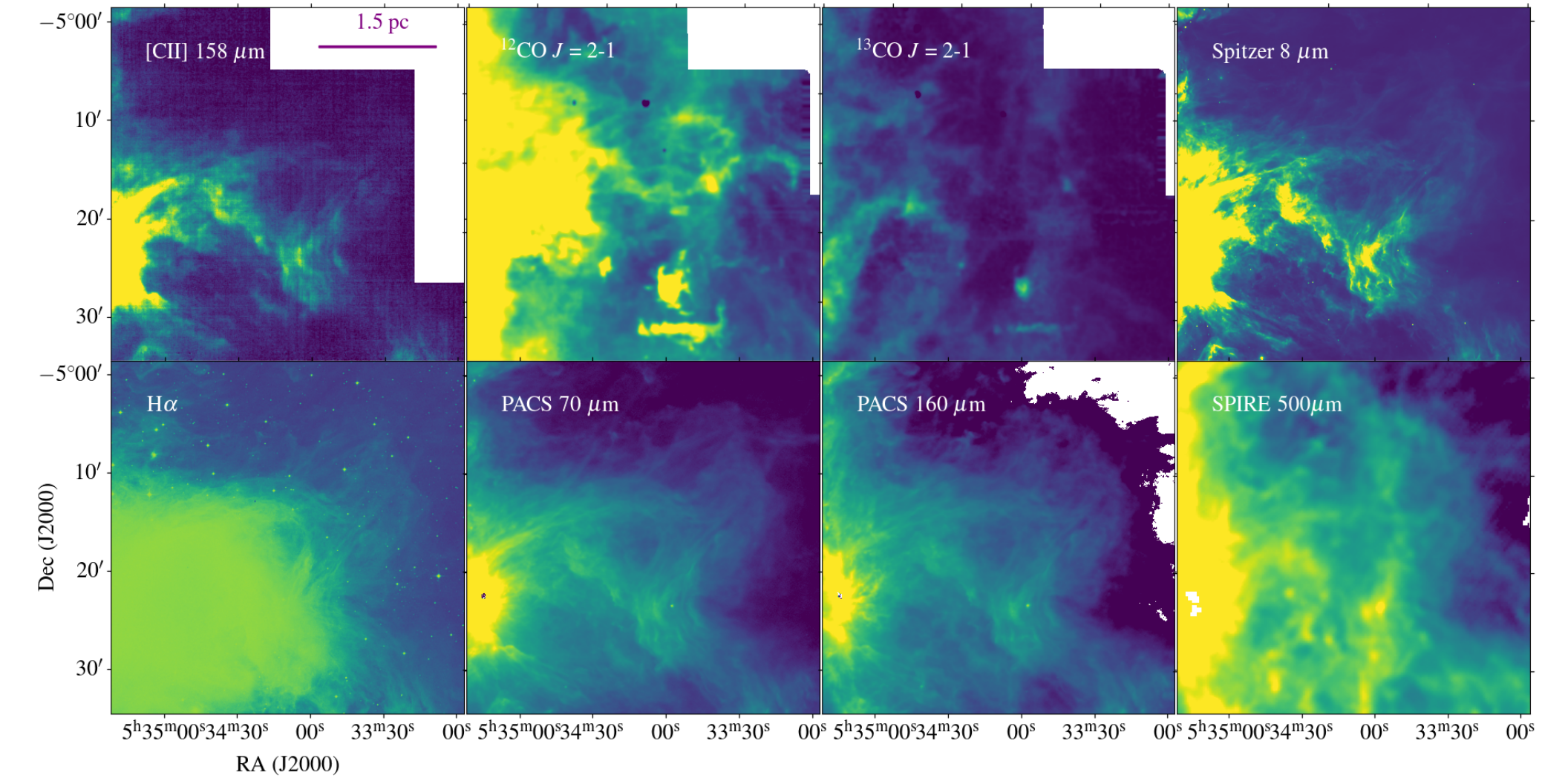} 
        \caption{Images of Orion's protrusion at different wavelengths and angular resolutions. The observed transition or frequency is given for each panel. \cii, $^{12}$CO (2-1), and $^{13}$CO (2-1) observations are integrated between $-$5 and $+14$~km~s$^{-1}$.}
    \label{fig:protsuion_gallery}
    \end{figure*}
    
    \subsection{Molecular Gas Observations} % CO (2-1)
    
    We use new $^{12}$CO \textit{J} = 2-1 (230.5~GHz) and $^{13}$CO \textit{J} = 2-1 (220.4~GHz) line maps taken with the IRAM~30m telescope. These data are part of the Large Program `Dynamic and Radiative Feedback of Massive Stars' (PI: J.~R.~Goicoechea). This project uses the old CO HERA and the new EMIR observations of the Orion Nebula. \citet{Goicoechea2020} describes how the old HERA and the new EMIR CO maps were merged. The last data relevant to this study were acquired during 2020. We extract the same region indicated in Fig.~\ref{fig:protsuion_gallery} from the original CO-cubes. The line intensities are presented in main-beam temperature ($T_\mathrm{mb}$) for both CO observations. In order to compare with the velocity-resolved \cii\,map, we smoothed the $^{12}$CO (2-1) and $^{13}$CO (2-1) data to the angular resolution of the SOFIA \cii\,maps of $16\arcsec$. The average rms noise level in these maps is 0.20~K in 0.41~km~s$^{-1}$ velocity channels. A more detailed description of the CO observations can be found in \citet{Goicoechea2020}.
    
    \subsection{Ionized Gas Observations}
    
    We use the H$\alpha$ images of the calibrated ESO/Digitized Sky Survey 2 (DSS-2) image obtained at the ESO/MPI 2.2-m telescope at La Silla \citep{DaRio2009}. The Orion Nebula has been observed on two different nights with the same observing strategy with 0.238$\arcsec$/pixel. After combining the dithered exposures, the final map has been created after trimming to the overlapping area. In the final map, the surroundings of the Trapezium stars are saturated but no saturation is seen in our region of interest. We extract the same region as indicated in Fig.~\ref{fig:protsuion_cii} to trace ionized gas with the H$\alpha$ map within the protrusion. The trimmed H$\alpha$ map we use is given in Fig.~\ref{fig:m42_Halpha}.
    
    \subsection{Far-IR photometric observations}
    
    We use the archival \textit{Herschel} images of the dust thermal emission for comparison to the \cii\,data, and in particular use this to estimate the mass of dust (and gas) associated with the protrusion. The Orion molecular clouds have been observed as part of the Gould Belt Survey \citep{Andre2010} in parallel mode using the Photoconductor Array Camera and Spectrometer \citep[PACS,][]{Griffin2010} and Spectral and Photo-metric Imaging Receiver \citep[SPIRE,][]{Poglitsch2010} instruments on-board \textit{Herschel}. We use the photometric images of PACS at 70~$\mu$m (beam FWHM of 5.6$\arcsec$), 100~$\mu$m (beam FWHM of 6.8$\arcsec$), and 160~$\mu$m (beam FWHM of 10.7$\arcsec$), and of SPIRE at 250~$\mu$m (beam FWHM of 18.1$\arcsec$) and 350~$\mu$m (beam FWHM of 25.2$\arcsec$). Because of the limited angular resolution, we refrain from using the longest wavelength SPIRE band at 500~$\mu$m in the comparison of the dust emission with the SOFIA \cii\,emission. Inspection of the 350~$\mu$m map reveals that omission of the 500~$\mu$m data does not compromise our analysis. We give more details about the model for fitting the \textit{Herschel} fluxes and the results of the spectral energy distribution (SED) fitting in Section~\ref{Section:Analysis}.
    
    A comparison between the \cii\,and \textit{Herschel} maps shows that the shorter wavelengths have almost the same morphology, which clearly represents FUV-heated warm dust in the protrusion (see Fig.~\ref{fig:protsuion_gallery}). However, faint emission, which could be physically connected to the protrusion itself, appears to the NW of the protrusion (see Fig.~\ref{fig:m42_schematic}). This component is also visible in most of the maps in Fig.~\ref{fig:protsuion_gallery}. Unfortunately, our \cii\,observations do not cover this component. 
    
    \subsection{Mid-IR Observations}
    
    We also make use of the Wide-field Infrared Survey Explorer (WISE) map of the Extended Orion Nebula\footnote{The WISE image of EON can be retrieved via: \url{http://wise.ssl.berkeley.edu/gallery_OrionNebula.html}} (EON; see also Fig.~\ref{fig:orion_wise}). Blue represents emission at 3.4~$\mu$m and cyan (blue-green) represents 4.6~$\mu$m, both of which come mainly from hot stars. Relatively cooler objects, such as the dust in the nebulae, appear green and red. Green represents 12~$\mu$m emission and red represents 22~$\mu$m emission. The field of view of the image is 3$^{\degr}$~$\times$~3$^{\degr}$ which covers the Veil and the extended emission coming from the dust. We trimmed the map to show a few striking jet-like structures that are present near the protrusion to the northeast of the Trapezium cluster.
    
    To trace the FUV-illuminated surface of PDRs, we use the \textit{Spitzer} 8~$\mu$m image (see Fig.~\ref{fig:protsuion_gallery}). The FWHM of the point spread function is 1.9$\arcsec$ at 8.0~$\mu$m. As in all observations, we extract the same region from the 8~$\mu$m image for further analysis.

%--------------------------------------------------------------------
\section{Results}\label{Sect:Results}

    Figure~\ref{fig:protsuion_gallery} shows the integrated intensity maps of the protrusion. The protrusion is clearly seen in \textit{Herschel} PACS 70 and 160~$\mu$m and SPIRE 500~$\mu$m images. We show three representative dust emission maps in Fig.~\ref{fig:protsuion_gallery} that trace the emission of dust heated by the Trapezium stars to $\sim$40~K. We also use the $^{12}$CO and $^{13}$CO $J$ = 2-1 observations to identify CO molecular gas exposed to intense FUV radiation. To confirm the location of PDRs, we overlay the \textit{Spitzer} 8~$\mu$m emission produced by PAHs on the \cii\,map in the right panel in Figure~\ref{fig:m42_8micron}. We see that the \cii\,emission has a similar distribution as the 8~$\mu$m emission map at the bottom and along the arm-like structure of the protrusion. We also compare the H${\alpha}$ emission with \cii\,to trace the ionized gas emission within the protrusion. The outlines of the protrusion are also quite apparent in H$\alpha$.
    
    \begin{figure*}[ht]
    \centering
        \begin{tabular}{cc}
        \begin{minipage}{0.99\textwidth} \centering
            \includegraphics[width=0.75\textwidth]{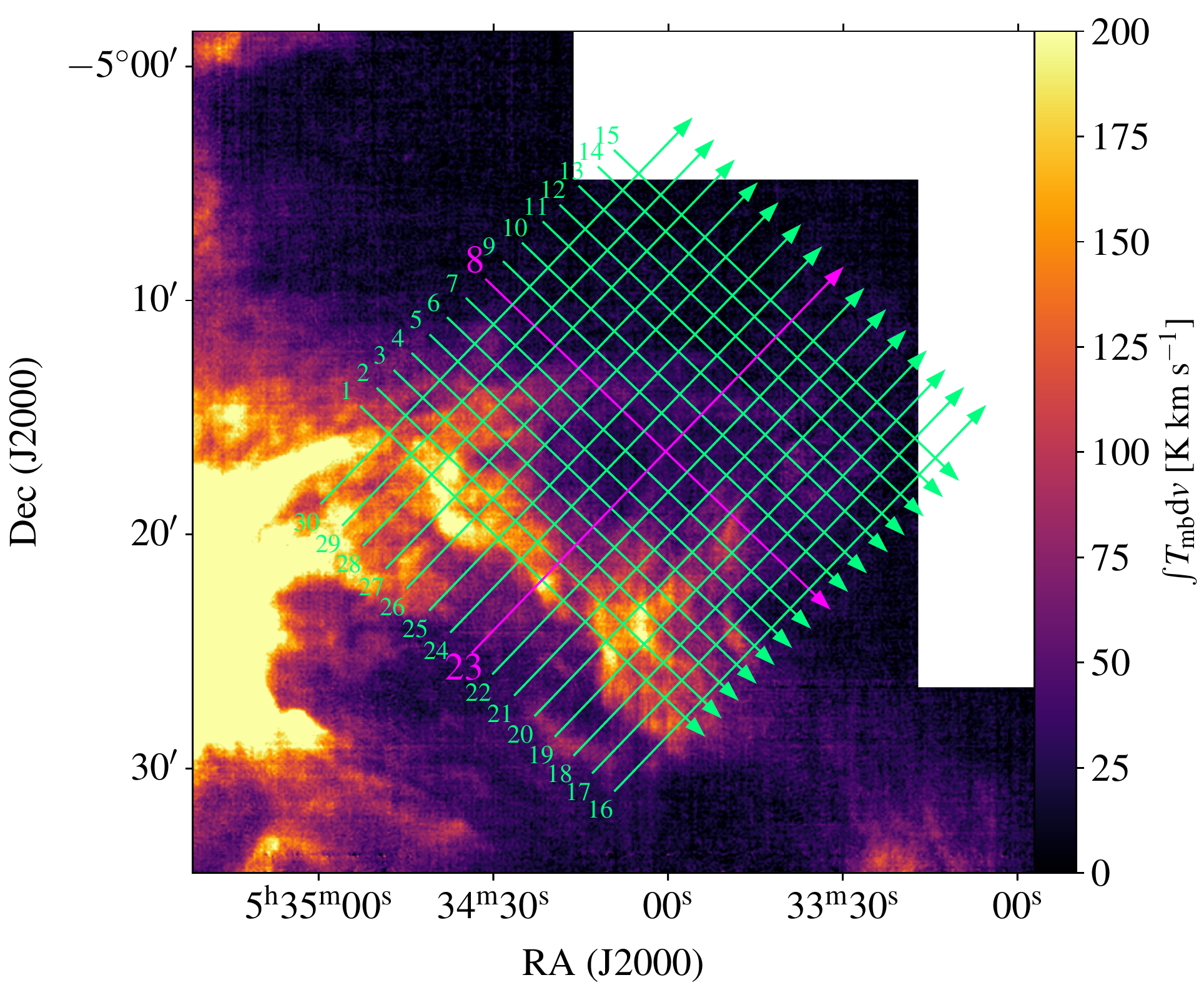} 
        \end{minipage} \\ 
        \begin{minipage}{0.99\textwidth}
            \includegraphics[width=0.5\textwidth]{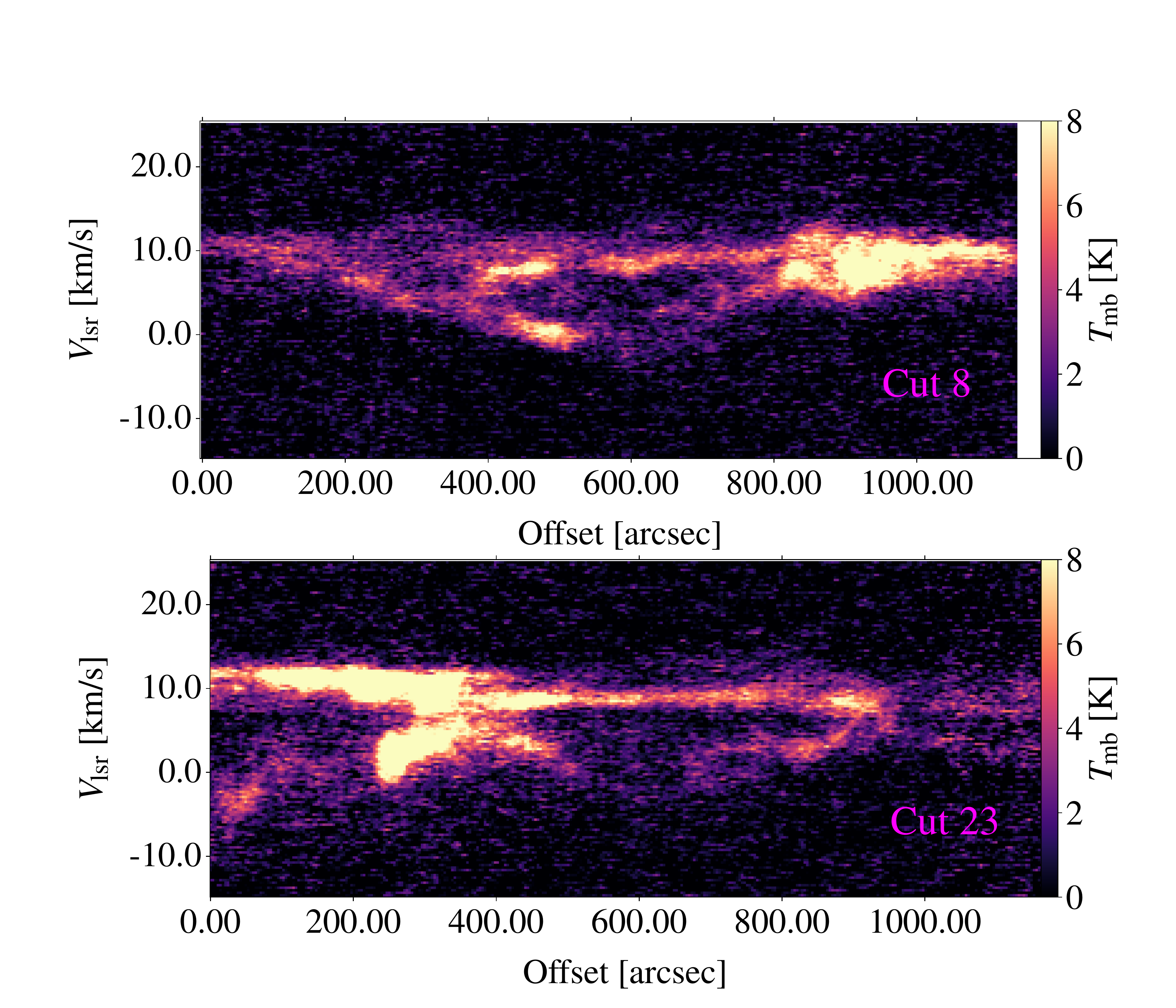}
            \includegraphics[width=0.5\textwidth]{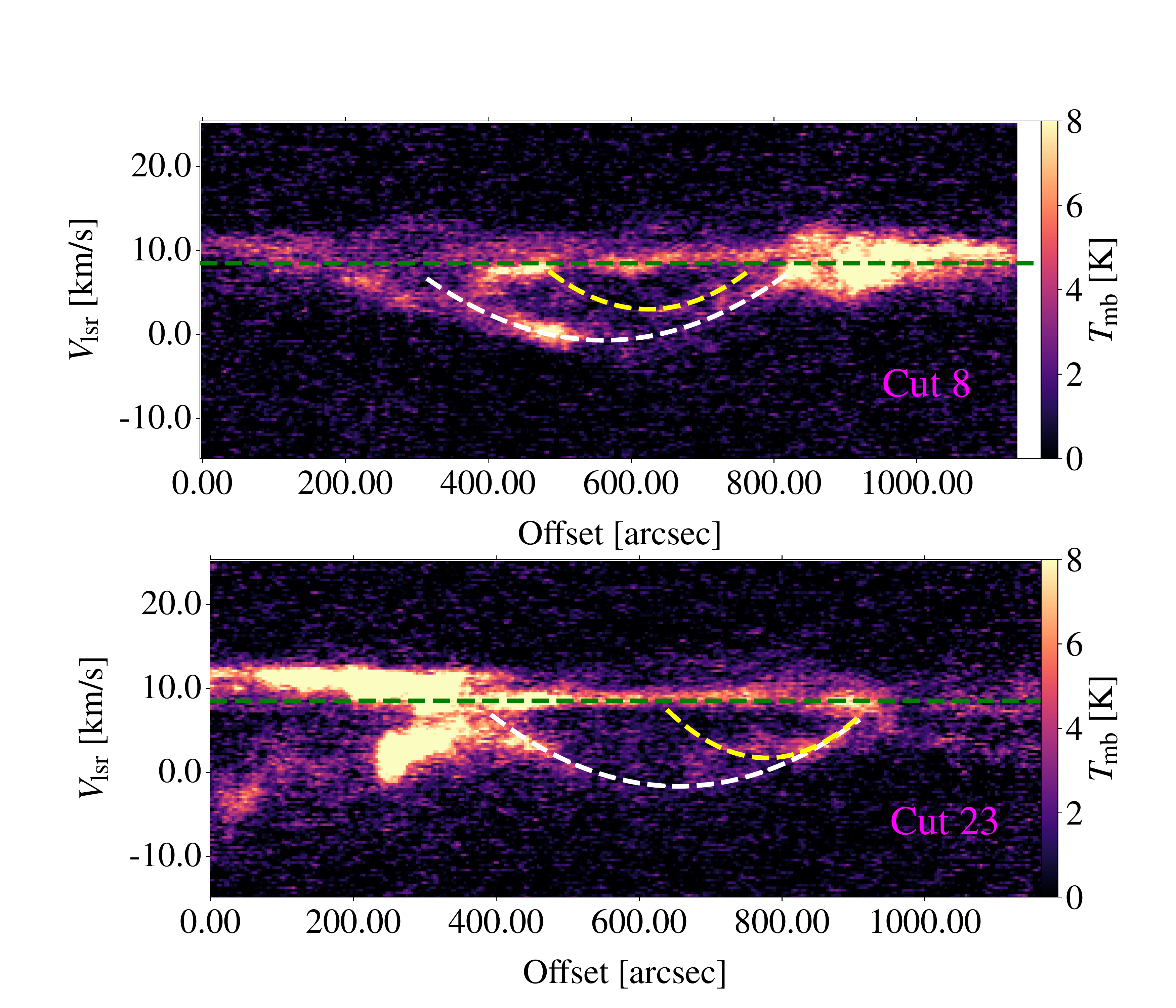}
        \end{minipage}
    \end{tabular}
        \caption{\textit{Top:} Selected crosscuts along the green arrows are overlaid on the integrated \cii\,intensity map. The number of the crosscuts is indicated at the starting point of the cut. \textit{Bottom:} The middle and bottom panels show the pv diagram generated along the magenta crosscuts (cuts 8 and 23, respectively). The pv diagram with horizontal green lines in both panels show the \cii\,emission produced by the FUV-illuminated surface of OMC and the arcuate white and yellow lines trace the shell expanding at 12~km~s$^{-1}$ and 6~km~s$^{-1}$, respectively. The remaining pv diagrams in Fig.~\ref{fig:allPVdiagrams1} and \ref{fig:allPVdiagrams2} have the same scale in both axes. A $^{12}$CO-PV diagram along the crosscut 23 is shown in Fig.~\ref{fig:COPVdiagrams} for comparison with \cii.}
    \label{fig:pv_diagrams}
    \end{figure*}

    Perusal of the individual channel maps (see Fig.~\ref{fig:PAH_cii_channelmaps3}) reveals that the protrusion is particularly noticeable in the local standard of rest (LSR) velocity range of $-$3 to $+$8~km~s$^{-1}$ in the \cii\,observations. It is clearly offset from the main \cii\,emission associated with the OMC$-$1 core at at $\gtrsim$ 9~km~s$^{-1}$. Unlike the \cii\,map, the protrusion does not appear in the $^{12}$CO $J$ = 2-1 velocity channel maps (e.g., see the $-$0.8~km~s$^{-1}$ channel map in Fig.~\ref{fig:PAH_cii_channelmaps3}) associated with the boundary of the Veil. On the other hand, $^{12}$CO $J$ = 2-1 shows a protrusion-like structure at higher velocities (12-13~km~s$^{-1}$) than those of the OMC$-$1 core (see Fig.~\ref{fig:PAH_cii_channelmaps3}). See Sect.~\ref{sec:lineprofiles} for more details about the origin of this high-velocity CO emission toward our protrusion.

\section{Analysis}\label{Section:Analysis}

    \subsection{Morphology of the protrusion}\label{sec:morphology}
    
    Our observations (see Fig.~\ref{fig:protsuion_gallery}) reveal expanding bow-shaped cavities in the northwest part of the Veil. The inside wall of these cavities is ionized, as shown by the H$\alpha$ emission, and the \cii, 8~$\mu$m, and 70~$\mu$m emission trace the surrounding PDR. First, we explore the protrusion itself, and later the ionizing star(s) and the origin of the protrusion. We fit the elliptical structure of the limb-brightened shell in the channel map at 12~km~s$^{-1}$ with a least-square fit to estimate the size and expansion velocity (see Sect.~\ref{sec:expansionvelocity}) of the protrusion. We find that the size of the protrusion is 1.3~$\pm$~0.1~pc from the Veil boundary to the NW direction. The minor and major axes of the model are 0.5~$\pm$~0.1 and 1.3~$\pm$~0.1~pc, respectively. The thickness of the shell we have derived is 0.1~$\pm$~0.05~pc. We assume an elliptical geometry to calculate the energetics of the protrusion in Sect.~\ref{sect:energy} because the channel maps suggest an elliptical morphology. 
    
    \begin{figure*}[!ht]
      \centering
       \includegraphics[width=0.7\hsize]{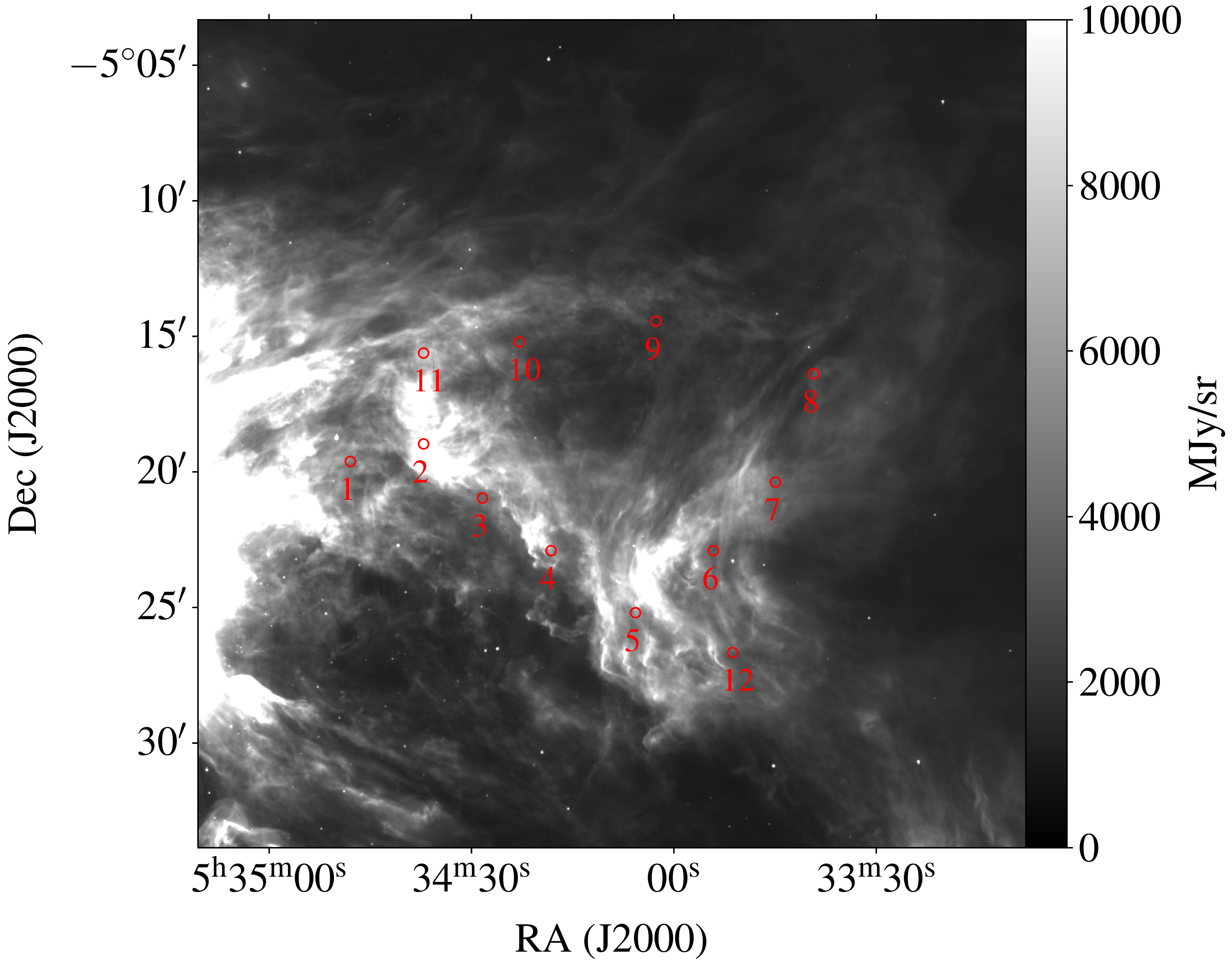}
        \includegraphics[width=0.99\hsize]{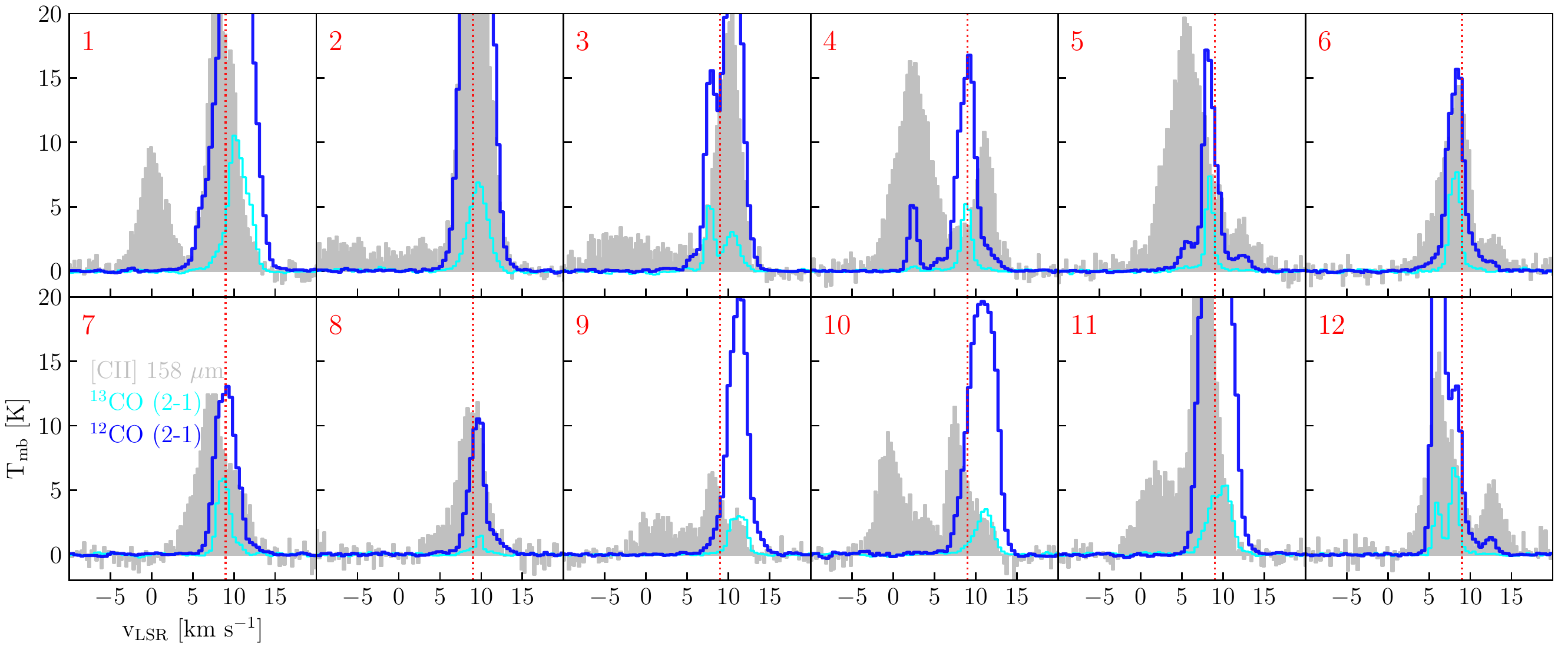}
    \caption{\textit{Upper panel:} \textit{Spitzer} 8~$\mu$m image of the protrusion. Red circles indicate twelve positions that we use to extract line profiles with an aperture of $16\arcsec$. \textit{Lower panel:} Velocity-resolved spectra of \cii\,(colored in gray), $^{12}$CO $J$~=~2-1 (blue), and $^{13}$CO $J$~=~2-1 (cyan) in the direction of protrusion for selected twelve positions in the upper panel. The vertical, red dotted line at 9~km~s$^{-1}$ marks the approximate velocity of the emission generated by the OMC and the associated star-forming molecular cloud behind the Veil.}
      \label{fig:lineprofiles}
    \end{figure*}

    We have examined the channel maps and determined the size of the expanding structure to be 1.3~pc in the southeast-northwest and 0.5~pc in the northeast-southwest direction. This ellipsoidal morphology is already quite apparent from the 8~$\mu$m and 70~$\mu$m dust emission maps. While morphologically, the structure resembles a half-cap in the plane of the sky, perusal of the pv diagrams shows that in all cross cuts, the structure starts and ends at the cloud velocity (+9~km~s$^{-1}$) even in the southeast-northwest direction (cf., cross cut 23 in Fig.~\ref{fig:pv_diagrams}). The observed PV diagrams are reasonably well fitted by a coherent half ellipsoidal shell.
    
    \begin{figure}[ht!]
      \centering
        \includegraphics[width=0.95\columnwidth]{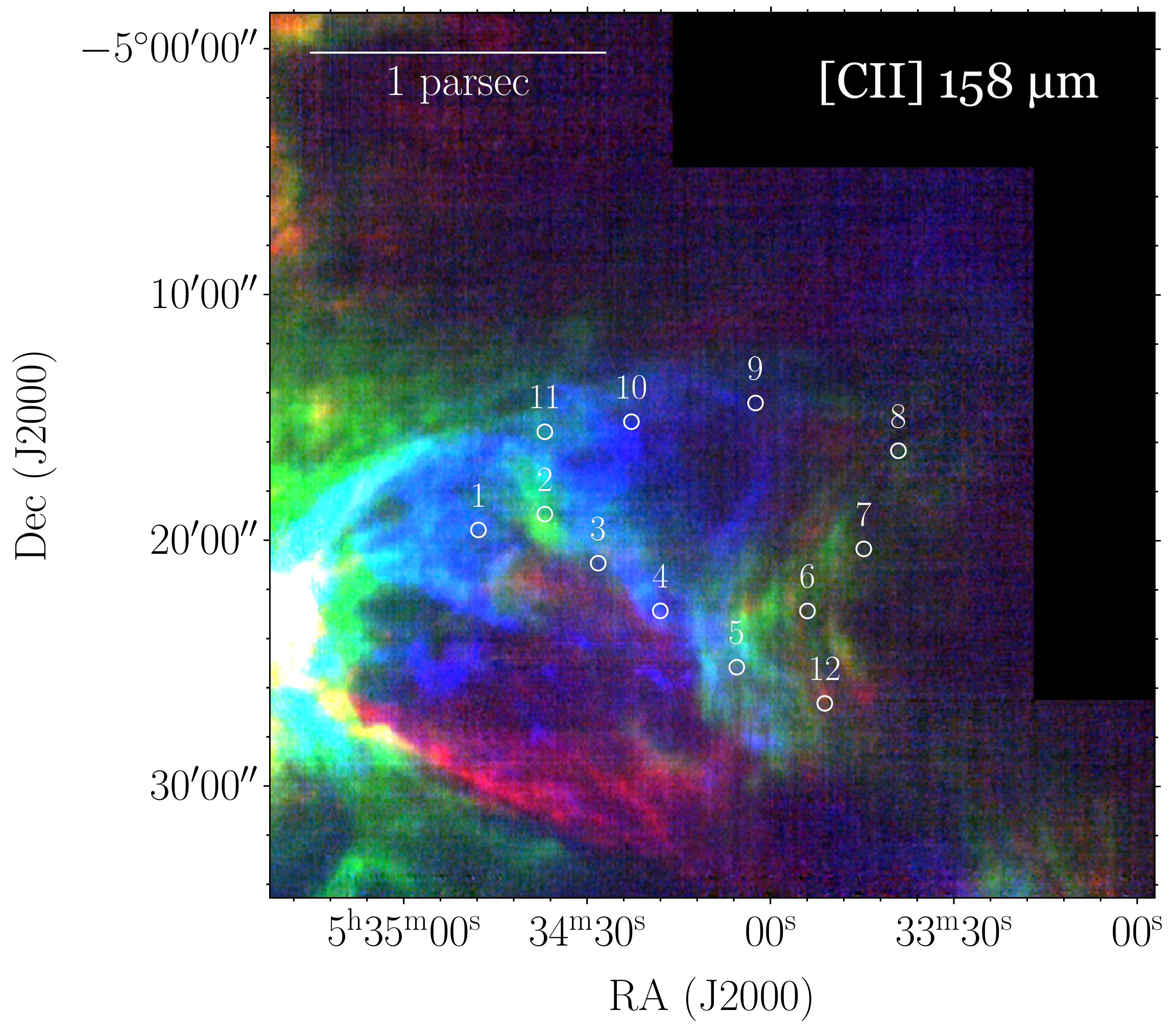}
        \includegraphics[width=0.95\columnwidth]{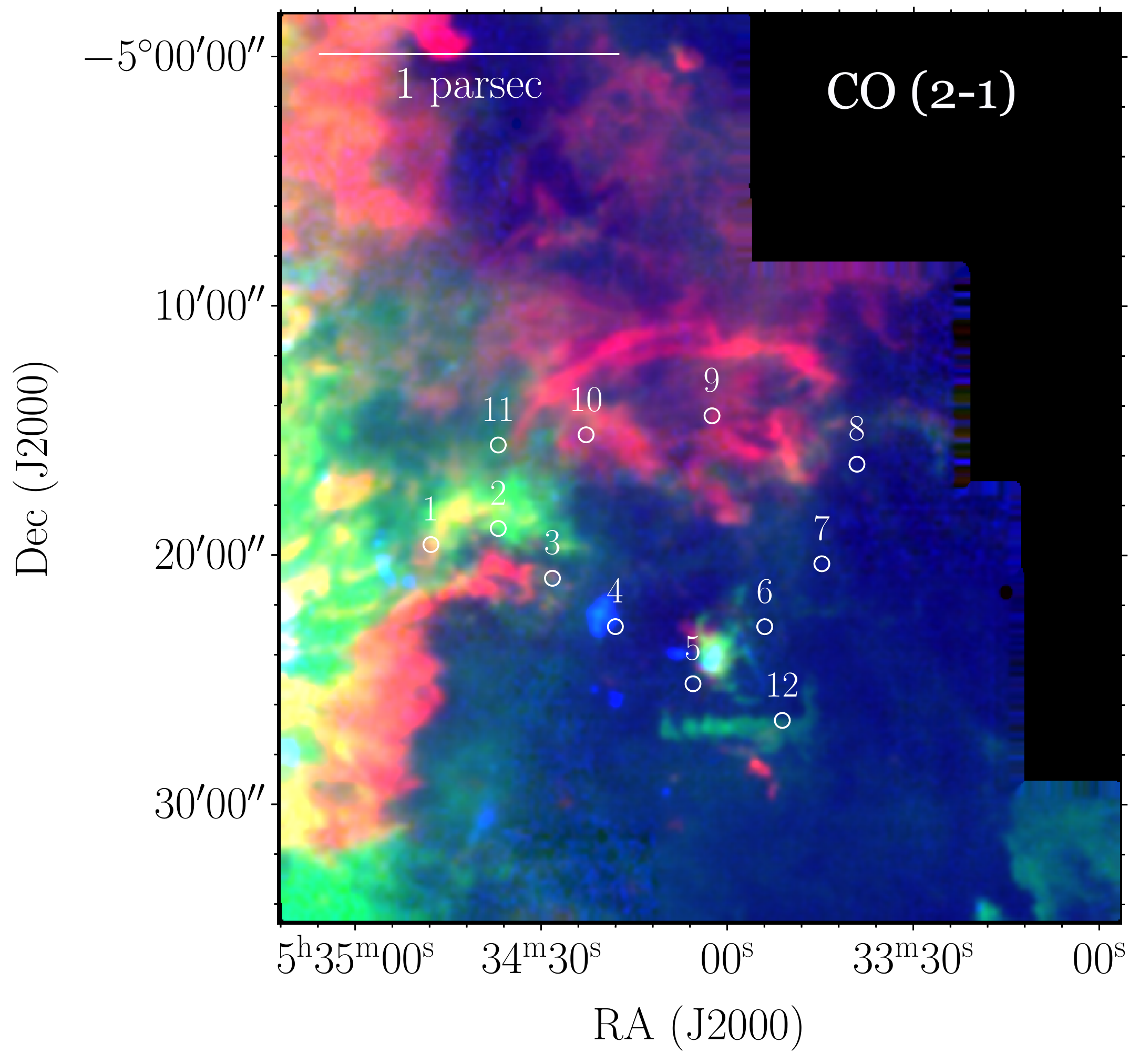}
        \caption{Three-color image of the protrusion created using three component estimated in \cii\,line profile in Fig.~\ref{fig:lineprofiles}. Blue emission is the integrated emission between $-$5 and $+$3~km~s$^{-1}$, green between $+$3 and $+$12~km~s$^{-1}$, and red between $+$12 and $+$15~km~s$^{-1}$ of the SOFIA \cii\,158~$\mu$m (upper panel) and IRAM~$^{12}$CO (lower panel) emission maps. White circles show the selected twelve positions in Fig.~\ref{fig:lineprofiles}.}
      \label{fig:m42_RGB}
    \end{figure}
    
%--------------------------------------------------------------------

    \subsection{Expansion velocity and timescale}\label{sec:expansionvelocity}
    
    Guided by the velocity channel maps, we quantify the characteristics of the protrusion in \cii\,position-velocity (pv) diagrams. We have created pv diagrams along thirty diagonal crosscuts, which are the 30$\arcsec$ wide white and magenta arrows in Fig.~\ref{fig:pv_diagrams}. We illustrate the results with two pv diagrams (cross cuts 8 and 23 in Fig.~\ref{fig:pv_diagrams}). The other pv diagrams are presented in Figs.~\ref{fig:allPVdiagrams1} and \ref{fig:allPVdiagrams2} and support the analysis presented here. Both pv diagrams in Fig.~\ref{fig:pv_diagrams} reveal two arc-like structures that are the tell tale signs of two half bubbles, both expanding only towards us.

    Inspection of all pv diagrams reveals two expanding shells. We fit these two arc-like structures in the pv diagram with a least-square fit over the chosen positions. The expansion velocity ($V_\mathrm{exp}$) of the first shell (yellow dashed line in Fig.~\ref{fig:pv_diagrams}) is $V_\mathrm{exp}$ = 6~$\pm$~0.2~km~s$^{-1}$ and the second (white dashed line in Fig.~\ref{fig:pv_diagrams}) $V_\mathrm{exp}$ = 12~$\pm$~0.2~~km~s$^{-1}$, which indicates the maximum expansion velocity of the outer shell. We fitted two pv diagrams (number 8 and 23) representing the maximum expansion of the protrusion using a simple bubble model (see Fig.~\ref{fig:pv_diagrams}). The emission at $V_\mathrm{LSR}$~=~+9~km~s$^{-1}$ (i.e. the green-dashed line) seen horizontally in both diagrams arises from the Orion cloud itself.
    
    When we take a closer look at the \cii\,channel maps in Fig.~\ref{fig:PAH_cii_channelmaps3}, we find two spatial components between $-$5 and $+$14~km~s$^{-1}$. The first component appears from $-$3 to $+$5~km~s$^{-1}$. The second component is identified between $+$6 and $+$14~km~s$^{-1}$ (see Fig.~\ref{fig:pv_diagrams}). We did not see the expanding shells in the CO channel maps and pv diagrams (see also Fig.~\ref{fig:COPVdiagrams}) and only detected an emission feature at $v_\mathrm{LSR}$ $+$2.7~km~s$^{-1}$ \citep[Globule \#10 of][]{Goicoechea2020}.
    
    The classical way to calculate the expansion timescale ($t_\mathrm{exp}$) for structures moving perpendicular to the line-of-sight is to use the ratio between the size of the outer shell and the maximum expansion velocity (size/$v_\mathrm{exp}$) \citep[see also][]{Beuther2002, Maud2015}. Above, we estimate the expansion velocity as 12~km~s$^{-1}$ using pv-diagram fit results. Using this expansion velocity and size (1.3~pc), $t_\mathrm{exp}$ we derive $\sim$1.1~$\times$~10$^5$~yr, which is $\sim$50\% of the expansion timescale of the entire Veil shell \citep{Pabst2019, Pabst2020}. 
    
    \subsection{Components of the protrusion}\label{sec:lineprofiles}
    
    In Figure~\ref{fig:lineprofiles}, we show the comparison of \cii\,158~$\mu$m, $^{12}$CO $J$=2-1, and $^{13}$CO $J$=2-1 spectra at twelve positions covering the protrusion to find sub-structures of the protrusion. $^{12}$CO and $^{13}$CO always have the similar profile, but at different brightness. The CO lines typically show two emission components (at +7 and +13~km~s$^{-1}$) at several positions (3, 4, 5, and 12) corresponding to the bottom of the protrusion. The velocity separation between the two CO peaks varies between 1$-$3~km~s$^{-1}$. These peaks in both CO isotopologues show small shifts (2$-$3~km~s$^{-1}$) to higher or lower velocities. The  absence of these velocity peaks in the \cii\,line profiles indicates that the CO emission is associated with structures deeper in OMC$-$1 that are not exposed to FUV radiation.
    
    In contrast, the \cii\,line shows a different behavior than CO, with the exception of position 2. In addition to the OMC, which emits predominantly at $V_\mathrm{LSR}$~=~$+$9~km~s$^{-1}$ (i.e., red-dotted line in Fig.~\ref{fig:lineprofiles}), we identify two other components on the \cii\,emission. To investigate the origin of these components, we have integrated the \cii\,emission between $-$5 and $+$3~km~s$^{-1}$ (blue is first component), $+$3 and $+$12~km~s$^{-1}$ (green is second component or OMC itself), and $+$12 and $+$15~km~s$^{-1}$ (red is third component). Note that the first component shifts to somewhat higher and lower velocities and that part of the profile of the first emission structure may be confused by emission of the OMC$-$1 core surface that dominates the total emission. We are therefore not able to use a fixed integration range for this component. The integration range is assumed based on positions~1 and 12 in Fig.~\ref{fig:lineprofiles}. Using integrated intensity maps, we create a three-color (or RGB) map of our protrusion using \cii\,and $^{12}$CO cubes which is shown in Fig.~\ref{fig:m42_RGB}. In the \cii\,RGB map, relative to the background OMC$-$1 core, the protrusion and the other structure are moving towards us at 9~km~s$^{-1}$. Together with the OMC, the blue component moving towards us is associated with the smaller (in size) expanding shell that we identified in the pv diagrams in Fig.~\ref{fig:allPVdiagrams1}. The presence of a red component at higher velocities (at 13~km~s$^{-1}$) than the OMC$-$1 core, suggests that there is a backward extension of the Veil shell. It is possible that the Veil shell on the rear side is tilted with respect to the background of the OMC$-$1 core, and sticking out of it, allowing for extension away from us. We do note though that the extension of the Orion Bar in the M42 \hii\,region is also quite prominent in this red channel and in that case, this velocity behavior could be related to complex morphology/velocity structures within the \hii\,region or at the PDR/\hii\,edges. In the $^{12}$CO RGB map, we track several components with velocities different from those of \cii. We conclude that the limb-brightened shell of the protrusion observed \cii\,does not contain CO and that the CO emission is associated with the molecular cloud in the background. 
    
    In addition, the cavity (redshifted emission in the vicinity of positions 8 to 11) has been identified with an expanding shell identified in CARMA CO $J$ = 1-0 observations by \citet{Feddersen2018}. They argued that Bruno 193 $-$ an F9IV star at the geometric center $-$ is driving this CO bubble. This bubble is thought to be embedded in the OMC$-$1 cloud behind the Veil. Based upon the kinematic information, we consider that this CO bubble is not related to the protrusion and this star is insufficient to ionize the gas; the more as this star is 7$\arcmin$ (0.85~pc) displaced from the center of our protrusion. The general morphology, the sub-components, and expanding shells are discussed in more detail in Section~\ref{sec:morphology}, \ref{sec:expansionvelocity}, and \ref{sec:lineprofiles}.

    \subsection{Kinetic energy and momentum}\label{sect:energy}

    To identify the driving mechanism of the protrusion, we calculate its momentum and kinetic energy. For this, we follow the same methods as in \citet{Pabst2020}. This also allows us to directly compare our results with the Veil shell \citep{Pabst2019}. To calculate the mass in the limb-brightened shell of the protrusion we use \textit{Herschel} PACS (70~$\mu$m, 100~$\mu$m, and 160~$\mu$m) and SPIRE (250~$\mu$m and 350~$\mu$m) maps. All maps are convolved to the SPIRE 350~$\mu$m beam size of $20\arcsec$ FWHM, as this resolution is comparable to the spatial resolution of SOFIA \cii. We convert the units of SPIRE maps from Jy~beam$^{-1}$ to Jy~px$^{-1}$ using the beam areas given in the HIPE\footnote{The software package for \textit{Herschel} Interactive Processing Environment (HIPE) is designed to work with the \textit{Herschel} data, including finding the data products, interactive analysis, plotting of data, and data manipulation.} manual. The flux densities at each pixel are modeled as a modified blackbody,
    \begin{equation*}
    I(\lambda) = B(\lambda, T_d)~\tau_0~\Bigg(\frac{\lambda_0}{\lambda}\Bigg)^\beta.
    \end{equation*}
    Here, $T_\mathrm{d}$ denotes the effective dust temperature, $\tau_0$ the dust optical depth at the reference wavelength $\lambda_0$, and $\beta$ the dust grain opacity index. The reference wavelength ($\lambda_0$) is 160~$\mu$m. $T_{d}$ and $\tau_{160}$ are free parameters. The dust emissivity index ($\beta$) is fixed at 2 in all models \citep{Goicoechea2015, Kavak2019, Pabst2019}. Maps of the fitted optical depth and dust temperature are shown in Fig.~\ref{fig:dust_temp_depth}. The statistical values of the dust temperature which are maximum, minimum, and median are 50~K, 20~K, and 26~K, respectively. The same statistics for the optical depth at 160~$\mu$m are 2~$\times$~10$^{-1}$, 8~$\times$~10$^{-4}$, and 2~$\times$~10$^{-3}$, respectively. Using an average value of the dust optical depth over the protrusion, we calculate the hydrogen column density:
    \begin{equation}
        N_H = \frac{\tau_{160}}{\kappa_{160} m_{H}} \simeq 6\times10^{24}\mathrm{cm^{-2}} \tau_{160}
    \end{equation}
    where $\kappa_{160}$ is the 160~$\mu$m dust opacity per H-atom\footnote{\url{https://www.astro.princeton.edu/~draine/dust/dustmix.html}} which is 2.3~$\times$~10$^{-25}$~cm$^2$/H-atom for $R_V$ = 5.5 \citep{Weingartner2001}. Using these values and the median optical depth, which is 2~$\times$~10$^{-3}$, we calculated the column density $N_H$~$\sim$~1.20~$\times$~10$^{22}$~cm$^{-2}$ (or a visual extinction of $A_\mathrm{v}$~=~8~mag). However, we also note that the limb-brightened shell of the protrusion seen in the \cii\,map does not appear in the $^{12}$CO $J$ = 2-1 map, indicating a low column density ($A_\mathrm{v}$~<~3~mag), in other words, a thin expanding shell. The high column density we derived reflects a difference in geometry. The dust emission estimate refers to the column density along the line of sight of a limb-brightened shell. Assuming a spherical homogeneous shell with a relative thickness of 0.1~pc, the column density estimates perpendicular to the surface of the protrusion will be about 0.2 times the observed column density and this is confirm the upper limit expected from the absence of CO in the protrusion.
    
    \begin{table}[!hb]
        \centering
        \begin{tabular}{r c c}
        \hline
        \hline
            & Veil Shell  & Protrusion  \\ 
        \hline
            size (pc)                                       & 2.7               & 1.3         \\
            thickness (pc)                                  & 0.5               & 0.1         \\
            density [$\times$~10$^3$~cm$^{-3}$]             & 1$-$10            & 0.1$-$1     \\
           %mass of ionized gas [$M_{\odot}$]               & 24                & XXX         \\
            $E_\mathrm{kin}$ [$10^{46}$~erg]                & 250               & 7           \\
            Momentum ($M_\odot$~km~s$^{-1}$)                & 20000             & 360$-$540   \\
           %$E_\mathrm{kin}$ of ionized gas [$10^{46}$~erg] & 6                 & XXX         \\
            expansion velocity [km~s$^{-1}$]                & 13                & 12          \\
            mass of neutral gas [$M_{\odot}$]               & 1500              & 30$-$45     \\
            %mass of ionized gas [$M_{\odot}$]              & 24                & 9           \\
            %electron density [cm$^{-3}$]                   &  XXX              & 120         \\
  %$\mathbb{N}_\mathrm{Lyc}$ [10$^{47}$~photons~s$^{-1}$]   & 70                & 5           \\
        \hline
        \end{tabular}
        \caption{Comparison of the masses and energetics of the protrusion with the Veil reported by \citet{Pabst2020}. The protrusion size is measured from the wall of the Veil shell to the outer shell in the NW direction.}\label{tab:mass_energetics}
    \end{table}

    \begin{figure*}[!h]
    \centering
        \includegraphics[width=\hsize]{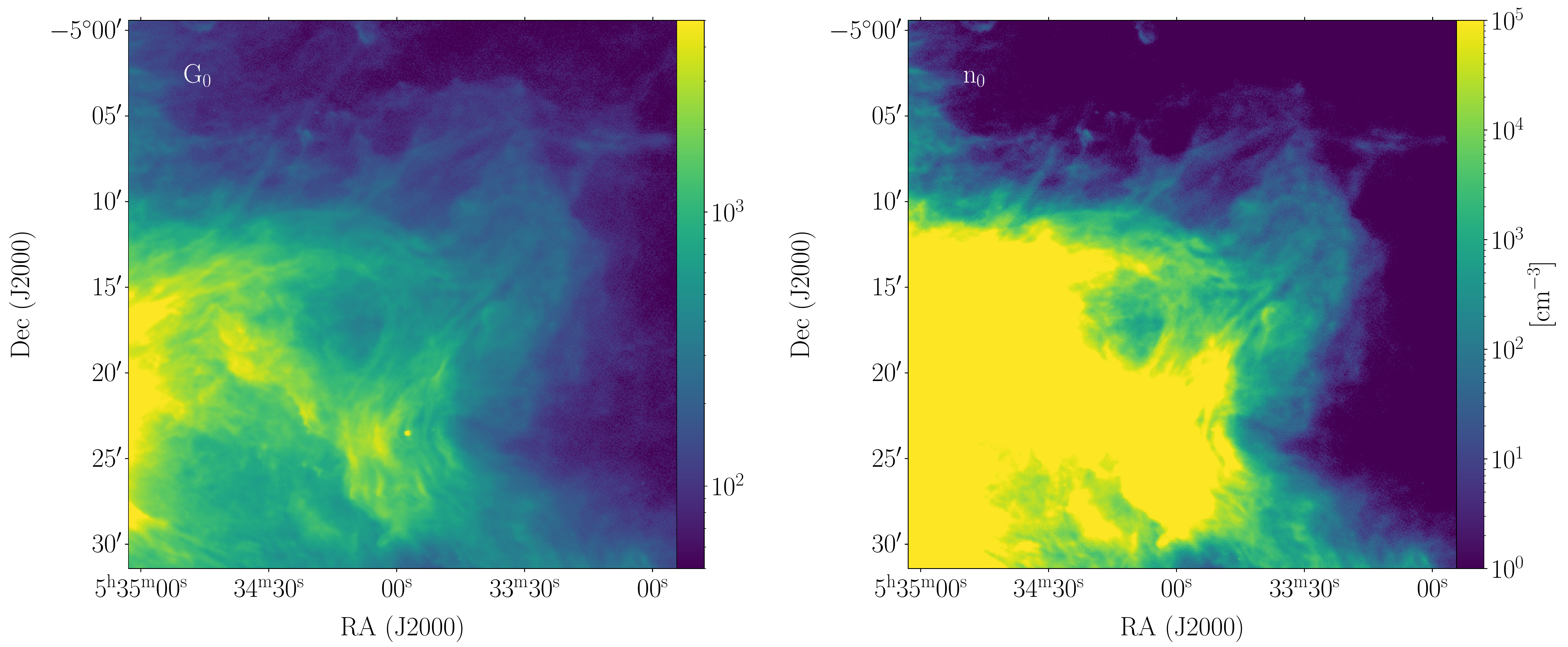}
        \caption{The map of the incident radiation field $G_0$ (left) and the density (right) of the protrusion for a face-on PDR model adopted from \citet{Tielens2010}. See Section~\ref{sect:energy} for a more detailed discussion.}
        \label{fig:G0andDensity}
    \end{figure*}
    
    Assuming elliptical geometry the mass of the limb brightened shell is given by the surface area, $S$, times the surface density along the line of sight; $M$~=~$S$~$N_\mathrm{H}$~$\mu$~$m_\mathrm{H}$. With the dimensions of the ellipse and a thickness of 0.1~pc, the surface area is calculated to be 0.13 pc$^2$, corresponding to a mass in the limb brightened shell of 18~$M_\odot$. A geometric correction factor (see Appendix~\ref{sect:geometriccorrection}) of 2.5 converts this then into the mass of the \cii\,emitting shell, $\sim$45~$M_\odot$. which is $\sim$3\% of the mass estimate of the Veil shell \citep[1500~$M_\odot$;][]{Pabst2020}. Using the mass estimate and the expansion velocity (12~km~s$^{-1}$), we calculate the kinetic energy ($E_\mathrm{kin}$) of the \cii\, gas tracing the neutral shell to be $\sim$7~$\times$~10$^{46}$~erg. Our energy estimate is $\sim$3\% of the kinetic energy of the entire expanding Veil shell \citep{Pabst2020}. Also, the momentum of the protrusion would be $\sim$540~$M_\odot$~km~s$^{-1}$.

    Additionally, we estimate the density of the protrusion by using the relation between $G_0$ and 70~$\mu$m reported by \citet{Goicoechea2020} for Orion to estimate the mass inside the protrusion. $G_0$ is given by, 
    \begin{equation} 
        \log_{10}(G_0) = (0.975\pm0.020)~\log_{10}(I_{70}) - (0.668\pm0.007)
    \end{equation}
    where $I_{70}$ is the 70~$\mu$m dust surface brightness in MJy~sr$^{-1}$. The median value of $G_0$ is $\sim$600 towards the protrusion, although a gradient can be seen in the $G_0$ map (see Fig.~\ref{fig:G0andDensity}). Using the estimate of $G_0$, we calculate the density of a face-on PDR using equation 9.4 of \citet{Tielens2010}, which is given in Eq.~\ref{eq:density}. We isolate the density and express it in terms of $G_0$.
    \begin{equation}\label{eq:density} 
    G_0 \simeq 10^2~\bigg(\frac{n_0}{10^3~\mathrm{cm^{-3}}}\bigg)^{4/3}
    \end{equation} 
    The resulting density map is also shown in Fig.~\ref{fig:G0andDensity}. The density decreases in the northwest direction from the boundary of the Veil to the outer shell of the protrusion. We can check our gas density from the observed \cii\,intensity using PDR models. For this purpose, we use the intensity of the \cii\,158~$\mu$m line emitted from the surface of an edge-on PDR as a function of the density and radiation field based on the PDR models\footnote{\url{http://dustem.astro.umd.edu/models/wk2006/cpweb.html}} of \citet{Kaufman1999} adopting an average $G_0$ of 600~Habings. This results in an average density of 10$^3$~cm$^{-3}$, in agreement with the estimates in Figure~\ref{fig:G0andDensity}. The density along the limb-brightened shell of the protrusion is comparable with that of the Veil shell \citep{Pabst2020} and two or three orders of magnitude lower than the Orion Bar \citep{Kavak2019, Pabst2020}. We calculate the mass of the shell of $\sim$30~$M_\odot$. This is in good agreement with the values calculated above in this section. Lastly, the momentum budget would be between 360 and 540~$M_\odot$~km~s$^{-1}$ (see Table~\ref{tab:mass_energetics} for a complete list of the characteristics of the protrusion).

    \subsection{Ionizing source}
    
    In Section~\ref{Sect:Results}, we show that the H$\alpha$ emission follows a similar morphology as the \cii\,emission. To understand the origin of ionized gas along the limb-brightened shell, we use the H$\alpha$ flux to estimate the source of the ionizing photons. We can make an estimate for the extinction associated with the protrusion from the thickness of the shell and the estimated column density of the limb brightened shell. Adopting a spherical half shell with a relative thickness of 0.1~pc, we estimate that the column density along the line of sight is 0.2 times the column density derived from the dust emission of the limb brightened shell, 2~$\times$~10$^{21}$ H~nuclei per cm$^2$. Using the extinction curve of \citet{Weingartner2001}, this corresponds to an extinction at H$\alpha$ of 1.1 mag. Correcting the observed surface brightness for extinction results in an intrinsic H$\alpha$ surface brightness of 525 MJy~sr$^{-1}$ or  2.7~$\times$~10$^{-7}$~erg~s$^{-1}$~cm$^{-2}$~arcsec$^{-2}$. We converted the surface brightness into emission measure\footnote{The emission measure is defined as $n_e^2$L, where $n_e$ is the electron density and L is the total path length in the ionized gas.} (EM) using Equation~\ref{eq:emissionmeasure}. Given a constant temperature of 8500~K obtained from radio recombination line observations \citep{Wilson1997},  
    \begin{equation}\label{eq:emissionmeasure}
        \left[\frac{EM}{\mathrm{pc~cm^{-6}}} \right] = 4.197~\times~10^{17}~\times~I_{H\alpha} 
    \end{equation}
    with H$\alpha$ in units of erg~s$^{-1}$~cm$^{-2}$~arcsec$^{-2}$. The EM we have derived as 1.40~$\times$~10$^{7}$~pc~cm$^{-6}$. We then calculated the total number of ionizing photons ($\mathbb{N}_\mathrm{Lyc}$) emitted by the star \citep[see Sect.~7.4.1 of][]{Tielens2010}. 
    \begin{equation}\label{eq:ionizingphotons}
        \mathbb{N}_\mathrm{Lyc} = A~\times~EM~\times~2.6~\times~10^{-13}
    \end{equation}
    where A is surface area in pc$^2$ and EM in pc~cm$^{-6}$. We find 1.8~$\times$~10$^{50}$~photons~s$^{-1}$. We measure the number of ionizing photons over a hole on the wall of the Veil of 1~pc$^2$, which is 1/16 of the total inner surface area of the Veil. In this case, the final number of ionizing photons is 1.1~$\times$~10$^{49}$~photons~s$^{-1}$. This indicates that the source of the ionizing photons should be an O-type star. The only O-star in the Trapezium cluster is $\theta^1$~Ori~C, the main ionizing star in Orion Nebula \citep{ODell2017}. Therefore, we conclude that the source of the ionized gas in the protrusion should be $\theta^1$~Ori~C.

\begin{figure}[t]
       \centering
        \includegraphics[width=\hsize]{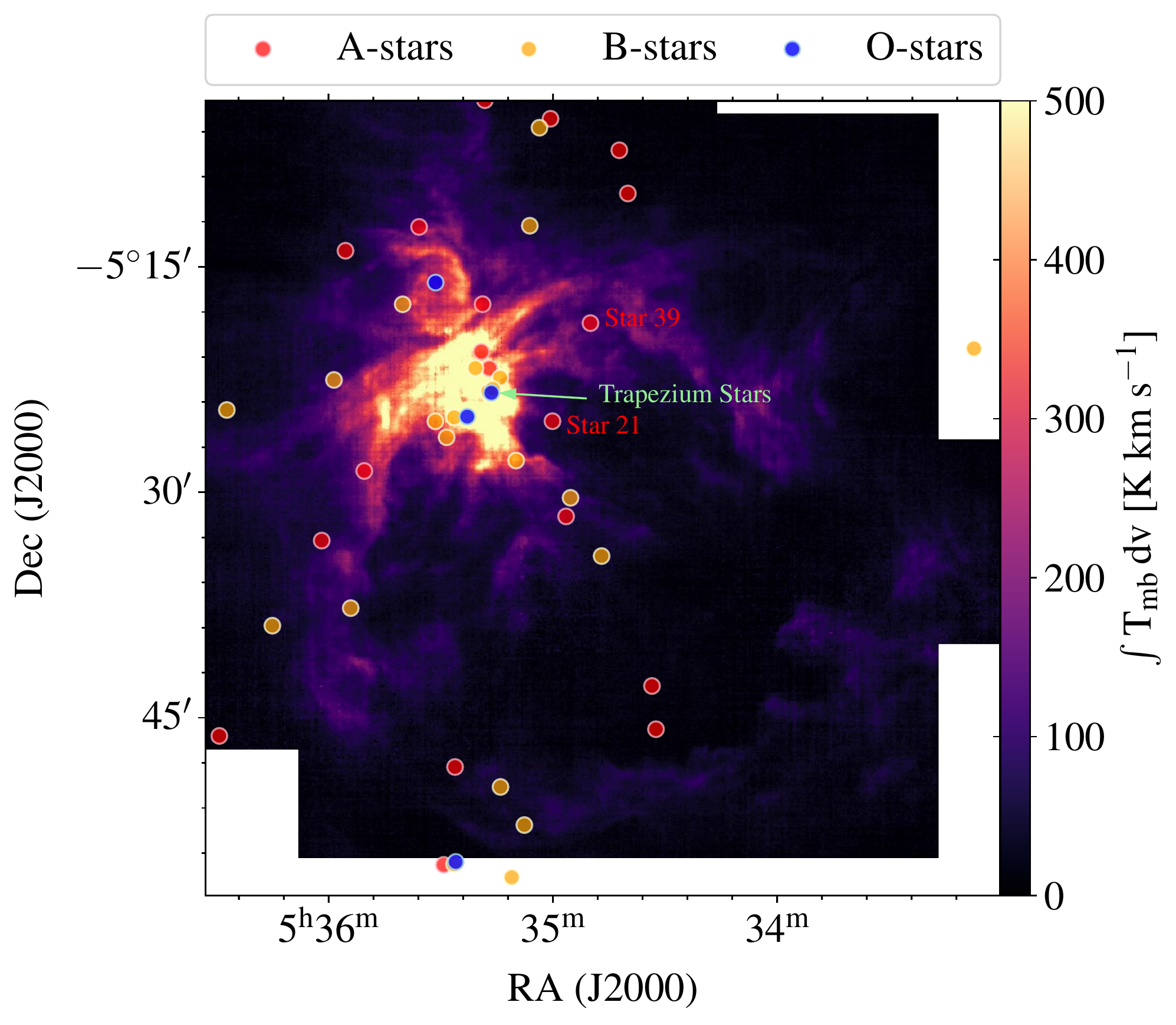}
        \caption{SOFIA \cii\,map of Orion with O$-$, B$-$, and A$-$stars found in SIMBAD. The list of stars retrieved from the archive is given in Table~\ref{tab:OBA_lists}. The blue, orange, and red circles are O$-$, B$-$, and A$-$stars, respectively. The light-green arrow indicate the positions of the Trapezium stars. Two A-stars (Star 21 and 39 in Table~\ref{tab:OBA_lists}) which are the nearest to the protrusion are also labelled.}
      \label{fig:cii_OBAstars}
    \end{figure}

    Additionally, to find other possible driving stars/sources in the protrusion, we display young stars and protostars identified with IRAC/\textit{Spitzer} \citep[green circles in Fig.~\ref{fig:m42_Halpha};][]{Megeath2005, Megeath2012}. In addition, we searched the SIMBAD archive for O$-$, B$-$, and A$-$stars within a 0.5$\arcmin$ circle around the Veil and listed 54 stars in Table~\ref{fig:cii_OBAstars} (see also Fig.~\ref{fig:cii_OBAstars}). This table consists of the ID and object name of the stars, coordinates in RA and Dec (in degree units), spectral type, and object type\footnote{For more information on object type, see \url{http://simbad.u-strasbg.fr/simbad/sim-display?data=otypes}}. The closest star to the protrusion is an A3 star (star 39 in Table~\ref{tab:OBA_lists})\footnote{In the GAIA DR2 survey, \citet{Grossschedel2018} reported a parallax of 2.4792~$\pm$~0.0374 mas corresponding to 404~$\pm$~6.1~pc indicating that Star 39 could be associated with the Orion Nebula.} which has a luminosity of 14~$L_\sun$ and a mass of about 2.0~$M_\sun$. We consider that this star is insufficient to ionize the surrounding gas and cause a protrusion because these type stars have a low effective temperature ($T_\mathrm{eff}$) and ionizing luminosity ($Q_\mathrm{i}$). Thus, we find no nearby powerful star that could ionize the gas or locally affect the shell or Veil in the north-west (see Section~\ref{sect:energy} for detailed analysis). Hence, the ionizing photons from the Trapezium cluster must be able to reach this surface almost unimpeded.

    \section{Discussion}\label{Section:Discussion}
    
    We examine three different scenarios in order to determine the driving mechanism of the protrusion. To begin, we discuss the role of the winds of the Trapezium stars, which are primarily responsible for the expansion of the whole Veil shell \citep{Pabst2019}. Later on, we examine the initial clumpiness of the pre-existing low-density regions \citep{Gudel2008}. Finally, we discuss the fossil outflow of Trapezium stars, which may provide directional feedback on the Veil shell, as seen in the massive star-forming region NGC~1333 \citep{Quillen2005}.
    
    \begin{figure*}[ht!]
       \centering
        \includegraphics[width=\hsize]{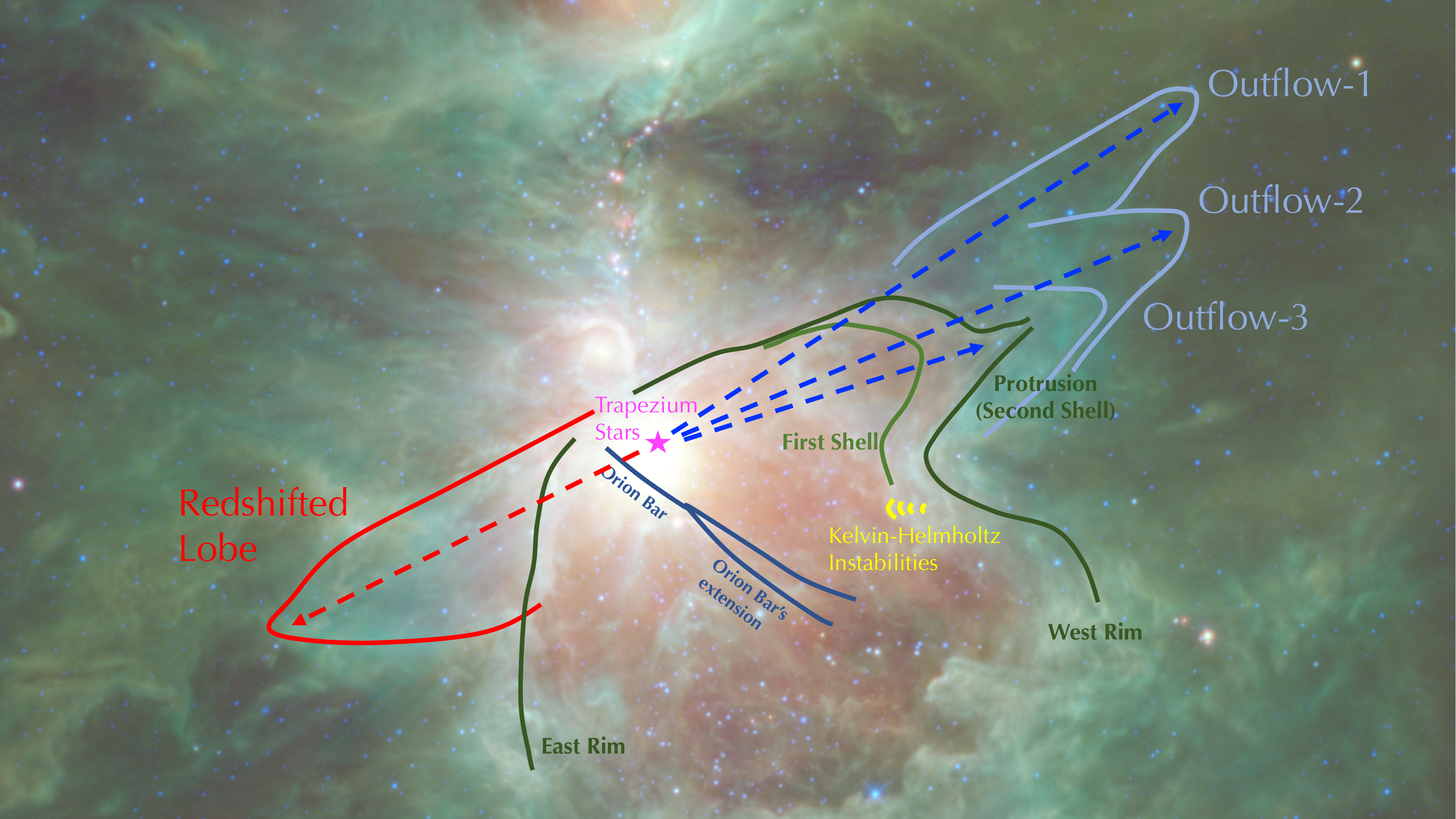} \\
        \caption{Schematic picture of the protrusion and the fossil bipolar outflow with apparent structures overlaid on the WISE image (see also Fig.~\ref{fig:orion_wise}). The suspected blue- and red-shifted outflows lobes are shown by blue and red colors, respectively. KH-instabilities (yellow) indicate Kelvin-Helmholtz instabilities reported by \citet{Berne2010}.}
      \label{fig:m42_schematic}
    \end{figure*}
    
    \subsection{Stellar winds}
    
    If the protrusion is driven by stellar winds of the Trapezium stars, in particular $\theta^1$~Ori~C, as found for the Veil \citep{Pabst2019}, the protrusion itself should expand like the Veil shell. However, despite that the velocity is (slightly) less than that of the Veil, the protrusion goes far beyond the Veil wall. Alternatively, the stellar winds could originate from another massive star within or near the protrusion. To check this, we superimpose the positions of the known O$-$, B$-$, and A$-$stars on the \cii\,map (Fig.~\ref{fig:cii_OBAstars}). There is no massive star within the protrusion. Only two A$-$stars are found near the protrusion. However, the nearest A$-$star (Star 39) does not follow the elongated morphology of the protrusion. The second A$-$star (Star 21 in Fig.~\ref{fig:cii_OBAstars}) is located at a comparable distance to the Trapezium stars. These findings force us to think of a pre-existing structure that is now being overtaken by the expanding Veil shell. 
    
    Another way to estimate the role of the winds is to compare them with X$-$ray observations, in which the hot X$-$ray emitting gas is traced inside the Veil. Using X$-$ray observations of the Veil, \citet{Gudel2008} showed that the X$-$ray emission from the ionized region indicates a hot plasma heated to a few 10$^{6}$~K by the shocks created by the stellar winds. In other words, the presence of X$-$ray emitting hot gas can be taken as an indication of stellar winds. However, there is no X$-$ray observation covering the protrusion. It should also be noted that X$-$ray emission is very susceptible to extinction by foreground material \citep{Gudel2008}. Therefore, X$-$ray observations may not be the best tool to investigate the effect of stellar winds, at least in our case. Imaging of optical line emission with the Apache Point Observatory (APO) will help us to detect the hot plasma (T > 30,000 Kelvin) inside the cavity (Bally et al., in prep).
    
    $\theta^1$~Ori~C drives the most powerful wind 2~$\times$~10$^{-7}$~$M_\sun$~yr$^{-1}$ with V$_\mathrm{wind}$ around 1000~km~s$^{-1}$. The winds of the lower-mass stars are weaker with lower mass-loss rates. As a rough estimate the total wind mass-loss rate from main-sequence OB stars is likely to be about 10$^{-6}$~$M_\sun$~yr$^{-1}$, or about 1~$M_\sun$ in 1~Myr. The momentum of the wind will be $\sim$1000~$M_\sun$~km~s$^{-1}$ in 1 Myr, comparable to that of the massive star outflows during their formation.
    
    \subsubsection{Lifetime of the protrusion}\label{sect:persistence}
    
    The protrusion has a limited lifetime due to the photo-ablation of its walls. Once the massive stars reach the ZAMS and begin to ionize their surroundings, photo-ablation of the inner walls of these cavities will start to fill their interiors with plasma. To first order, the plasma will expand at the speed of sound in ionized gas at $V_\mathrm{[CII]}$~=~10~km~s$^{-1}$. Using our mass estimations in Table~\ref{tab:mass_energetics}, the surface area of the protrusion (0.385~pc$^2$), and the incident flux of Lyman continuum photons (2~$\times$~10$^{49}$ s$^{-1}$ for $\theta^1$~Ori~C), we can estimate the mass-loss rate of the protrusion walls and how long the walls would survive ($t_\mathrm{sur}$). The mass-loss rate is given by,
    
    \begin{equation}
        \frac{dM}{dt} = f~\mu~m_H~n_e~V_\mathrm{[CII]}~R^2
    \end{equation}
    where $f$ is a factor of order unity depending on geometry which is taken to be $\sqrt{3}$ to recover the Str\"omgren condition for a spherical HII region. The plasma density ($n_e$) can be calculated assuming that the incident Lyman continuum flux (L(LyC)/(4~$\pi$~D$^2$)) equals the recombination along a path length ($R$),
    \begin{equation}\label{eq:plasmadensity}
        n_e = \frac{f}{D}~\Big[\frac{L(LyC)}{4~\pi~\alpha_B~R}\Big]^{0.5}
    \end{equation}
    where $\alpha_B$ is the Case B recombination coefficient of H; 2.6~$\times$~10$^{-13}$~cm$^3$~s$^{-1}$. The number of electron-proton recombinations per unit volume and unit time is equal to $n_e$~$n_p$~$\alpha_B$. Using Eq.~\ref{eq:plasmadensity}, we derive a plasma density ($n_e$) of $\sim$2~$\times$~10$^3$~cm$^{-3}$. The mass-loss rate from the protrusion walls is 1.8~$\times$~10$^{-4}$~$M_\odot$~yr$^{-1}$. Therefore, the lifetime of the protrusion (i.e., $t_\mathrm{sur}$ = M/(dM/dt), where $M$ is the mass of \cii\,emitting protrusion walls) is $\sim$1.6~$\times$~10$^5$~years, which is consistent with the age of the Trapezium stars and the expansion timescale derived in Sect.~\ref{sec:expansionvelocity}, but not with the age of the O9 to early B-stars below the bright Orion Bar (the $\theta^2$~Ori~A stars) whose age is older than 10$^6$~years. We argue that the location of the protrusion is an ideal place to break Orion's Veil and ventilate its hot plasma before a possible supernova occurs ($\sim$5~$\times$~10$^6$~years; see also \citet{Williams1997}).
    
    \subsection{Blow-out of the Veil shell}
    
    The wind of $\theta^1$~Ori~C would produce a spherical bubble only if there were no obstacles blocking the propagation of the wind and the post-shock hot plasma. However, we know that there is a dense cloud in the region of the K-H instabilities (see Figs.~\ref{fig:protsuion_gallery} and \ref{fig:m42_schematic}) containing CO \citep{Berne2010} whose surface is affected by radiative feedback (maybe a wind) from the Trapezium cluster. The H-alpha ionization front of the nebula wraps around this structure. An even larger protrusion is visible south of this obstacle, especially in wide-field \sii\,image (Fig.~\ref{fig:orion_sii}). The core of this obstacle is seen in CO; it is the feature marked as the "KH Ripples" in the figures in CARMA Orion CO survey \citep[][]{Kong2018}. This also prevents hot plasma from flowing westward and forces it to head south of the Veil \citep{Gudel2008, Bally2010}. A possible model for the northwest protrusion is that the plasma driving the Veil shell has found a path of least resistance towards the northwest. Moreover, the south of the obstacle of the protrusion is seen at $V_\mathrm{LSR}$~$+$5 to $+$8~km~s$^{-1}$ (see the $^{12}$CO-RGB map in Fig.~\ref{fig:m42_RGB}). Hence, blow-out of the nebula around this obstacle is certainly possible\footnote{See the channel movie to see how the western rim of the Veil shell changes morphologically around this obstacle: \url{https://owncloud.sron.nl/owncloud/index.php/s/s8utQcp4M84AbCY}.}. However, both the lower expansion of the protrusion than the Veil shell and the bipolar jet-like structures seen toward the Veil shell depicted in Fig.~\ref{fig:m42_schematic} are difficult to reconcile with this scenario.

    \subsection{Fossil bipolar outflow}

    The slightly lower expansion velocity of the protrusion than the Veil and its extension beyond the boundary of the Veil suggest that the protrusion is a pre-existing structure in the OMC$-$1 core that is now being overtaken by the Veil bubble. Following Bally et al. (in prep), we suggest that this pre-existing structure is the result of fossil outflow activity in the OMC$-$1 core created during the accretion phase of the massive protostars in the Trapezium cluster. In Fig.~\ref{fig:m42_schematic}, we show probable blue- and red-shifted outflows on the WISE image. Blue-shifted outflows ejected from one or more of the Trapezium stars (Outflow-1, Outflow-2, and Outflow-3) create our protrusion in the northwestern part of the Veil shell. If there is truly a bipolar fossil outflow, we may expect to observe a red-shifted lobe towards the eastern rim of the Veil shell. The suspected red-shifted lobe is also associated with high-velocity $^{12}$CO emission (see Fig.\ref{fig:orion_wise}). Additionally, the red-shifted lobe appears to have broken the eastern rim, ejecting a tail-like structure at the head of the outflow (see Fig.~\ref{fig:orion_wise}).
    
    \begin{figure}[!ht]
      \centering
        \includegraphics[width=\hsize]{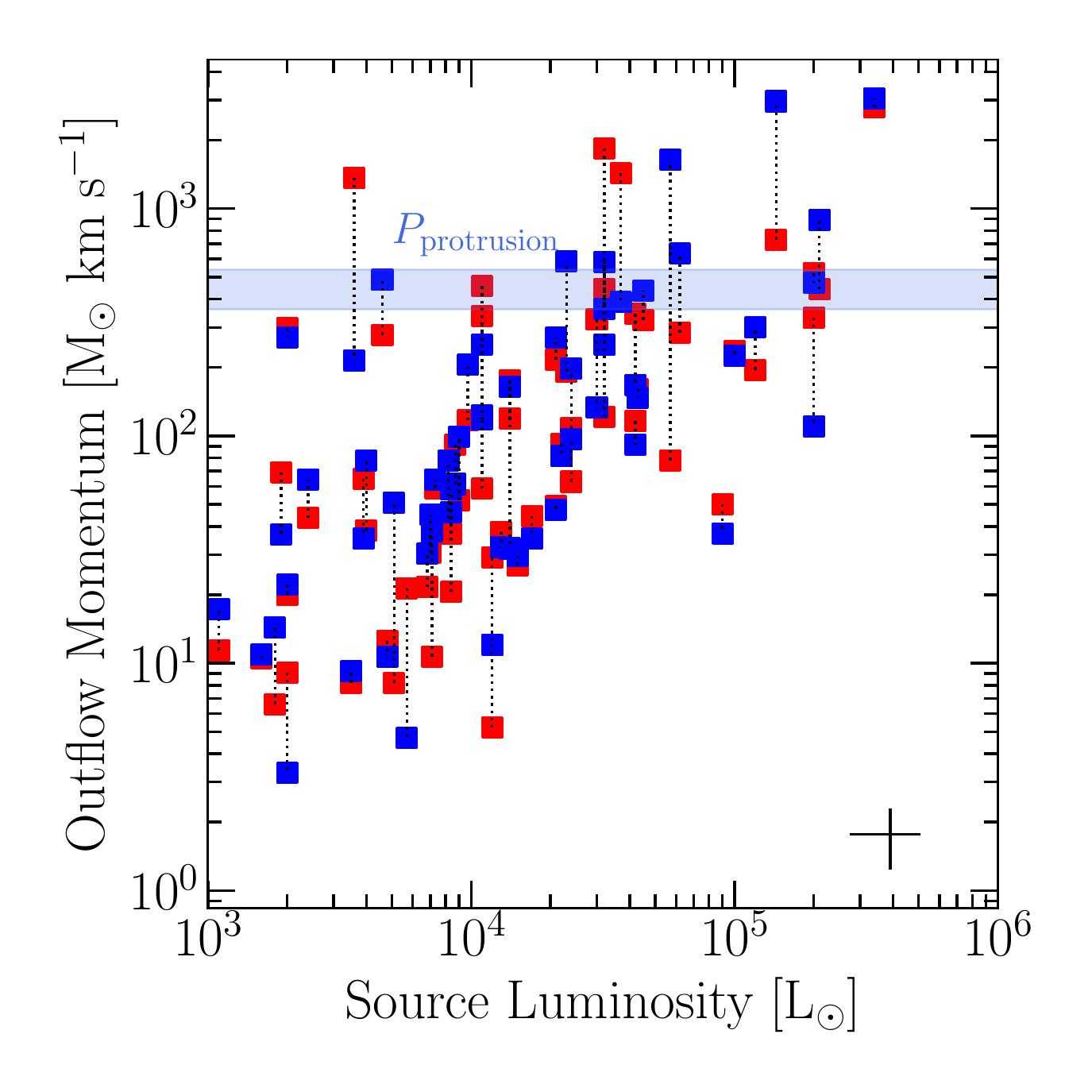}
        \caption{Momentum of outflows from massive young stellar objects as a function of the source luminosity of the cores \citep{Maud2015}. The blue and red symbols indicate the blue- and red-shifted outflow lobe values, respectively which are joined by a dotted line for each source. The horizontal blue-shaded range indicates the momentum of the protrusion, which is between 360$-$540~$M_\odot$~km~s$^{-1}$. The cross at the bottom-right shows the uncertainty for both axes.}
      \label{fig:outflow_mass}
    \end{figure}
    
    Once the protostellar jet switches off, the cavity blown by this jet will enter the momentum conserving phase and expand while slowing down. As $\theta^1$~Ori~C entered its main sequence phase, its stellar wind started to blow the Veil bubble. The large amount of momentum involved in this kinematic structure could indicate outflow activity associated with the formation of the most massive star. To identify the possible protostellar source(s), we use the bolometric source luminosity and momentum of the outflows from \citet{Maud2015}. The momentum of $M_\mathrm{outflow}$ red- and blue-shifted lobes are given individually with red- and blue-shifted squares in Fig.~\ref{fig:outflow_mass}, respectively. The interpretation of the relation in Fig.~\ref{fig:outflow_mass} is that the jet or wind from the most luminous protostar drives the strongest and most powerful outflows. For the scatter in the momentum values, \citet{Maud2015} argue that it is caused either by outflow inclination angles or by multiple outflows driven by sources within dense cores. We also emphasize that this type of outflow activity is generally found in systems with ages less than a few times 10$^4$~yr \citep{Arce2007} and hence is a clear signature of protostellar activity.
    
    Using the relation in Fig.~\ref{fig:outflow_mass}, we estimate that a massive dense shell with a momentum of $\sim$540~$M_\odot$~km~s$^{-1}$ would require a luminosity of 3~$\times$~10$^4$ to 3~$\times$~10$^5$~$L_\odot$. This corresponds to B0 to O7 type stars \citep[cf. for stellar parameters of O and B stars][]{Vacca1996} and several stars in the Trapezium region could be responsible. Likely, $\theta^1$~Ori~C, the most massive star, is the culprit. We do notice that there are several other jet-like morphological structures present in the 8~$\mu$m and WISE images in the area of the protrusion (Fig.~\ref{fig:m42_schematic} and \ref{fig:orion_wise}). Our \cii\,observations do not cover these structures and therefore we have no kinematic information on their expansion. Further (deeper) studies are warranted to determine their kinematics. Here, we recognize that these structures may indicate the presence of multiple protostellar outflows for example associated with the several of the Trapezium star cluster. Alternatively, these jet-like structures may reflect intermittent activity of a single, precessing object in a binary of the Trapezium cluster. We do note that the trajectories of these jet-like structures trace back to the Trapezium stars (Fig.~\ref{fig:m42_schematic}).
    
    At this point, we speculate that the protrusion was likely created by outflow activity when accretion in a protostar-disk structure was accompanied by a jet/wind in the polar directions. On this basis, it can be argued that the Trapezium stars (specifically $\theta^1$~Ori~C) should have formed via disk-mediated accretion. This model of massive star formation is supported by recent studies which have found disks \citep{Cesaroni2017}, outflows \citep{LopezSepulcre2010, SanchezMonge2013}, and jets \citep{Sanna2018, Kavak2019}. If the protrusion is made of fossil outflow cavities, there have to be the counter flows corresponding to the red-shifted lobe of the northwest protrusions from the Trapezium cluster. WISE and 8~$\mu$m images show a vague protrusion in the opposite direction of the northwestern protrusion. Given the blue-shifts of the NW protrusions, this component should be the red-shifted lobe (i.e., the red arrow shows the red-shifted lobe in Fig.~\ref{fig:m42_schematic}). However, the \cii\,emission is weak at this position preventing us to study this red-shifted lobe in this work. Note also that the fossil outflow activity is not related to the explosive outflow and the H$_2$ fingers seen in near-IR lines \citep{Bally2017}, as these fingers are still far ($\sim$1.5~pc) from the boundary of the Veil shell and relatively recent ($\sim$500 yr) `explosive' event.
    
    \section{Conclusion}\label{Section:BreakingVeil}
    
    In this study, we investigate the origin of the protrusion in the northwestern part of the Orion Veil shell using velocity-resolved \cii\,158~$\mu$m observations. The protrusion, which appears as a half elliptical cap on the Veil, expands at a velocity of 12~km~s$^{-1}$ with a radius of 1.3~pc. In Sect.~\ref{Section:Discussion}, we examine three possible mechanisms which could drive the protrusion on the northwest. We propose that the protrusion is formed by extinct or previously active Trapezium star outflows. During the early stages of massive star formation, the outflows of the Trapezium stars weaken the northwest portion of the Veil. Later on, as the massive stars reach the ZAMS, their radiation ionizes the surrounding gas while their stellar wind starts to blow a bubble filled with hot gas. The EUV and FUV photons can travel barely unimpeded in the fossil outflow cavity, illuminating the protrusion and lighten it up in H$\alpha$, PAH emission, and \cii\,emission. This suggests that mechanical feedback is the responsible mechanism for the formation of the protrusion rather than radiative feedback. Moreover, in Section~\ref{Sect:Results}, we also see that the lack of CO detections in the protrusion indicate a low $N_H$, or in other words, a thin shell in the northwestern Veil. In Sect.~\ref{sect:persistence}, we also show that the fossil outflow activity could cause breaks in the ionization front of the Orion Veil because of photo-ablation from the protrusion walls, making the protrusion a suitable place for the Veil to break up.
    
    Moreover, the diagonal pv diagrams parallel to the direction of expansion, in particular cuts 18, 19, and 20 in Fig.~\ref{fig:pv_diagrams}, show \cii\,emission that extends somewhat beyond the protrusion. The densities of the limb-brightened shell are lower (a factor of up to two) at the head of the protrusion. Outflows, particularly Outflow-3 in Fig.~\ref{fig:m42_schematic}, appear to be associated with the chimney-like top of the protrusion, suggesting that the Veil shell has already been pierced here. This location could be a suitable place for the bubble to break. Furthermore, beyond the area mapped in \cii, the outflows and their extended morphology are also seen in the \textit{Spitzer}~8~$\mu$m image, the dust emission maps of \textit{Herschel}~PACS~70~$\mu$m and WISE observations. Thus, future more sensitive \cii\,observations could clarify whether or not the Veil is already broken at the location of the protrusion.
    
    If the protrusion is formed by a bipolar fossil outflow, the south-east cavity carved by the red-shifted lobe of the fossil outflow (see Fig.~\ref{fig:m42_schematic}) should exhibit similar characteristics as our protrusion. However, unlike our protrusion, the south-east cavity is not filled with ionized hydrogen plasma, but is associated with high-velocity CO emission. We require further data to determine what protects it from Lyman continuum illumination from the Trapezium or $\theta^2$~Ori~A. For example, a shock tracer such as the H$_2$ v = 2-1 S(1)/v=1-0 S(1) line intensity ratio (at 2.24~$\mu$m and 2.12~$\mu$m, respectively) and/or \feii\,1.644~$\mu$m (e.g., observed with Keck or JWST in the near future) tracing the presence of shocked gas toward the south-east protrusion may reveal the shocked gas within the red-shifted lobe of fossil outflow. Additionally, the emission measure of the south-east protrusion could be too low. If this is true, we should expect to witness free-free plasma at the cavity walls. This, however, is not the case. Thus, further effort should be directed toward determining the nature of the protrusion carved by the red-shifted lobe and establishing the involvement of bipolar fossil outflows in the formation of the north-west protrusion.
    
    In summary, three mechanisms can contribute to the formation, morphology, and expansion of \hii\,regions: (i) pre-cavitation caused by powerful bipolar outflows; (ii) ionization and thermal expansion of the \hii\,region in the early phases; and (iii) stellar winds in the late phases. The protrusions can readily explained by pre-cavitation caused by collimated bipolar outflow. However, we cannot rule out the possibility of obstacle molecular clouds playing a role. Another research may focus on the outflow kinematics beyond the Veil's NW protrusion and the obstacles at the ankle of the protrusion.
    
\begin{acknowledgements}
    We want to thank Martin Vogelaar (Groningen) for his help for solving Python programming problems and Anthony G.A. Brown (Leiden) for retrieving the list of O$-$, B$-$, and A$-$ stars from the SIMBAD database. We also thank Marc William Pound and Mark Wolfire for their help on the PDR Toolbox. Studies of interstellar dust and gas at Leiden Observatory are supported by a Spinoza award from the Dutch Science agency, NWO. JRG thanks the Spanish MICINN for funding support under grant PID2019-106110GB-I00 partially based on IRAM 30m telescope observations. IRAM is supported by INSU/CNRS (France), MPG (Germany), and IGN (Spain). This study was based on observations made with the NASA/DLR Stratospheric Observatory for Infrared Astronomy (SOFIA). SOFIA is jointly operated by the Universities Space Research Association Inc. (USRA), under NASA contract NAS2-97001, and the Deutsches SOFIA Institut (DSI), under DLR contract 50 OK 0901 to the University of Stuttgart. upGREAT is a development by the MPI für Radioastronomie and the KOSMA/Universität zu Köln, in cooperation with the DLR Institut für Optische Sensorsysteme. We acknowledge the work, during the C+ upGREAT square degree survey of Orion, of the USRA and NASA staff of the Armstrong Flight Research Center in Palmdale, the Ames Research Center in Mountain View (California), and the Deutsches SOFIA Institut.
\end{acknowledgements}

\bibliographystyle{aa}
\bibliography{aanda.bib}

\begin{appendix}

    \section{Geometric correction Factor}\label{sect:geometriccorrection}
    
\begin{figure}[!h]
        \centering
        \includegraphics[width = \columnwidth]{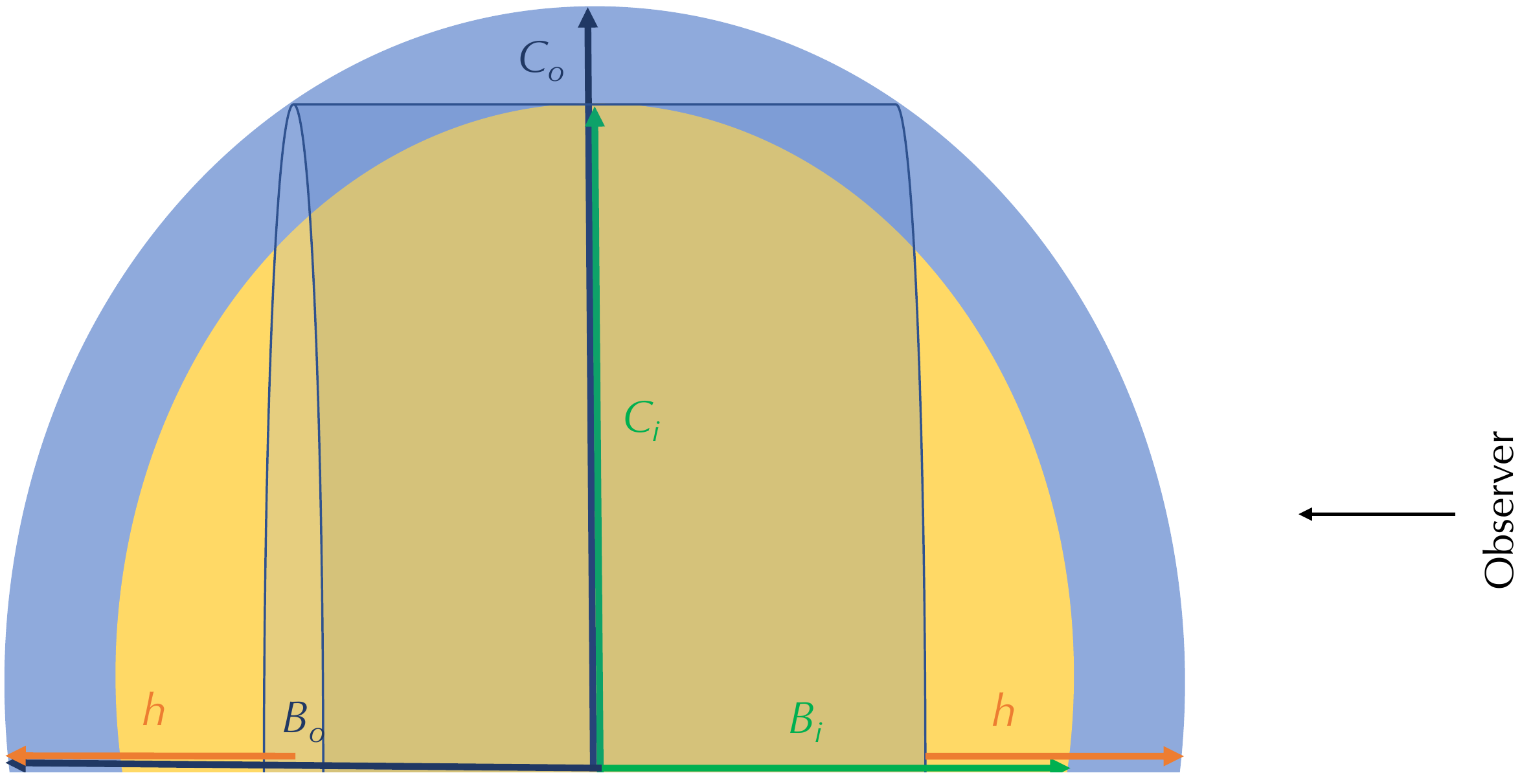}
        \caption{Shell geometry}
        \label{fig:shellgeometry}
    \end{figure}

    The limb-brightened shell observed in different tracers is seen as an arc of emission. If we assume that the emission seen in the dust tracer, the \cii\,line or the CO line is proportional to the total volume, then we need some geometry to figure out what the enhancement factor, $f_\mathrm{v}$, is that scales the volume of the limb brightened part to that of the full shell. We consider two concentric nested ellipsoids with major diameter 2$C_o$ and 2$C_i$ and minor diameter 2$B_o$ and 2$B_i$. The protrusion is half of this ellipsoid (see Fig.~\ref{fig:shellgeometry}). If the cap height is $h$, the cap volume is given by;
    
    \begin{equation}
        V_\mathrm{cap} = \frac{\pi}{3}~C_o^2~\Big(\frac{h}{B_\mathrm{o}}\Big)^2~(3B_o-h)
    \end{equation}
    
    The base surface area of the cap is,
    \begin{equation}
        A_\mathrm{cap} = \pi~(C_o~B_o)~\Big(\frac{h}{B_\mathrm{o}}\Big)~(2-\frac{h}{B_o})
    \end{equation}
    
    The volume of the cylinder is,
    \begin{equation}
        V_\mathrm{cyl} = 2~(B_o-h)~A_\mathrm{cap}
    \end{equation}
    
    The total volume of the outer ellipsoid is, 
    \begin{equation}
        V_\mathrm{ell} = \Big(\frac{4\pi}{3}\Big)~B_o^2~C_o
    \end{equation}   
    
    The volume of the rim is then,
    \begin{equation}
        V_\mathrm{rim} = V_\mathrm{ell}-2~V_\mathrm{cap}-V_\mathrm{cyl}
    \end{equation}

    We compare this to the volume in between the two nested ellipsoids, 
    
    \begin{equation}
        V = \frac{4\pi}{3}\Big(B_o^2~C_o-B_i^2~C_i\Big)
    \end{equation}
    
    Actually, as the ellipsoid only protrudes half out of the Veil, we should divide all of these volumes by two. As we are really interested in $V_\mathrm{rim}$/$V$ these factors two drop. Now we have to express h in the sizes of the inner and outer ellipsoid. The base area of the cap is equal to the surface area of the inner spheroid.
    \begin{equation}
        A_\mathrm{cap} = \pi~B_i~C_i
    \end{equation}
    Thus, h can be found from,
    \begin{equation}
        A_\mathrm{cap} = \pi~(C_o~B_o)~(\frac{h}{B_o})~(2-\frac{h}{B_o}) = \pi~B_i~C_i
    \end{equation}

    For $B_\mathrm{o}$~=~0.5~pc, $C_\mathrm{o}$~=~1.3~pc, $B_\mathrm{i}$~=~0.4~pc, and $C_\mathrm{i}$~=~1.2~pc, we find that the height of the cap would be 0.244~pc. Using this, we estimate that the volume of the \cii\,emitting limb-brightened rim is 2.5 of the total volume of half ellipse in Fig~\ref{fig:shellgeometry}. In this case, the mass in the limb-brightened shell would be between 45~$M_\odot$ which is in good agreement with the mass estimation of 30~$M_\odot$ based on the PDR models. Finally, the mass of the limb-brightened shell is between 30$-$45~$M_\odot$.
    
    \section{Additional Maps}
    
\begin{figure}[ht!]
       \centering
        \includegraphics[width=\columnwidth]{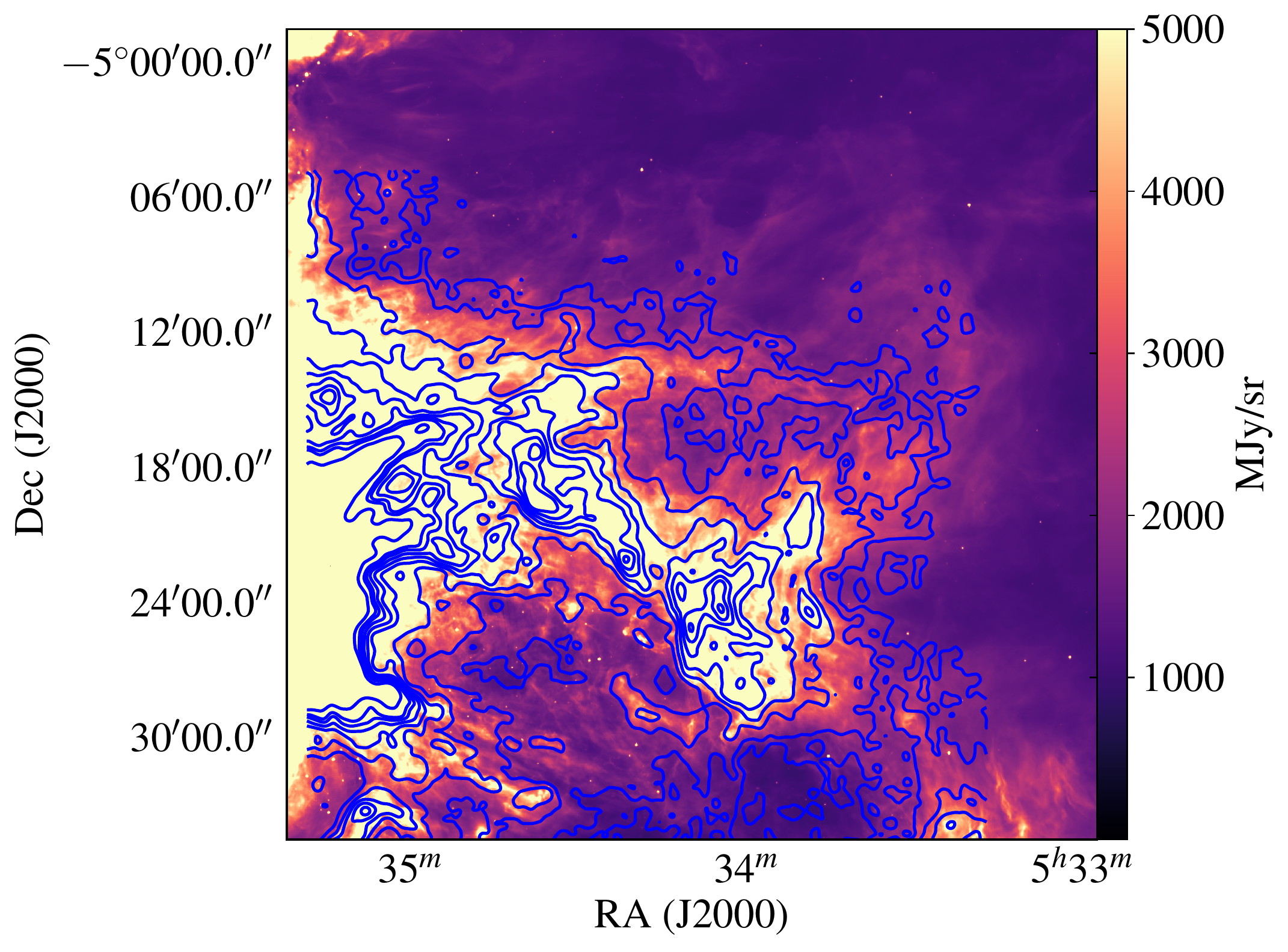} \\
        \caption{\textit{Spitzer}~8~$\mu$m image, which outlines the PDR surfaces. The blue contours show the SOFIA \cii\,line integrated emission.}
      \label{fig:m42_8micron}
    \end{figure}

\begin{figure}[ht!]
       \centering
        \includegraphics[width=\columnwidth]{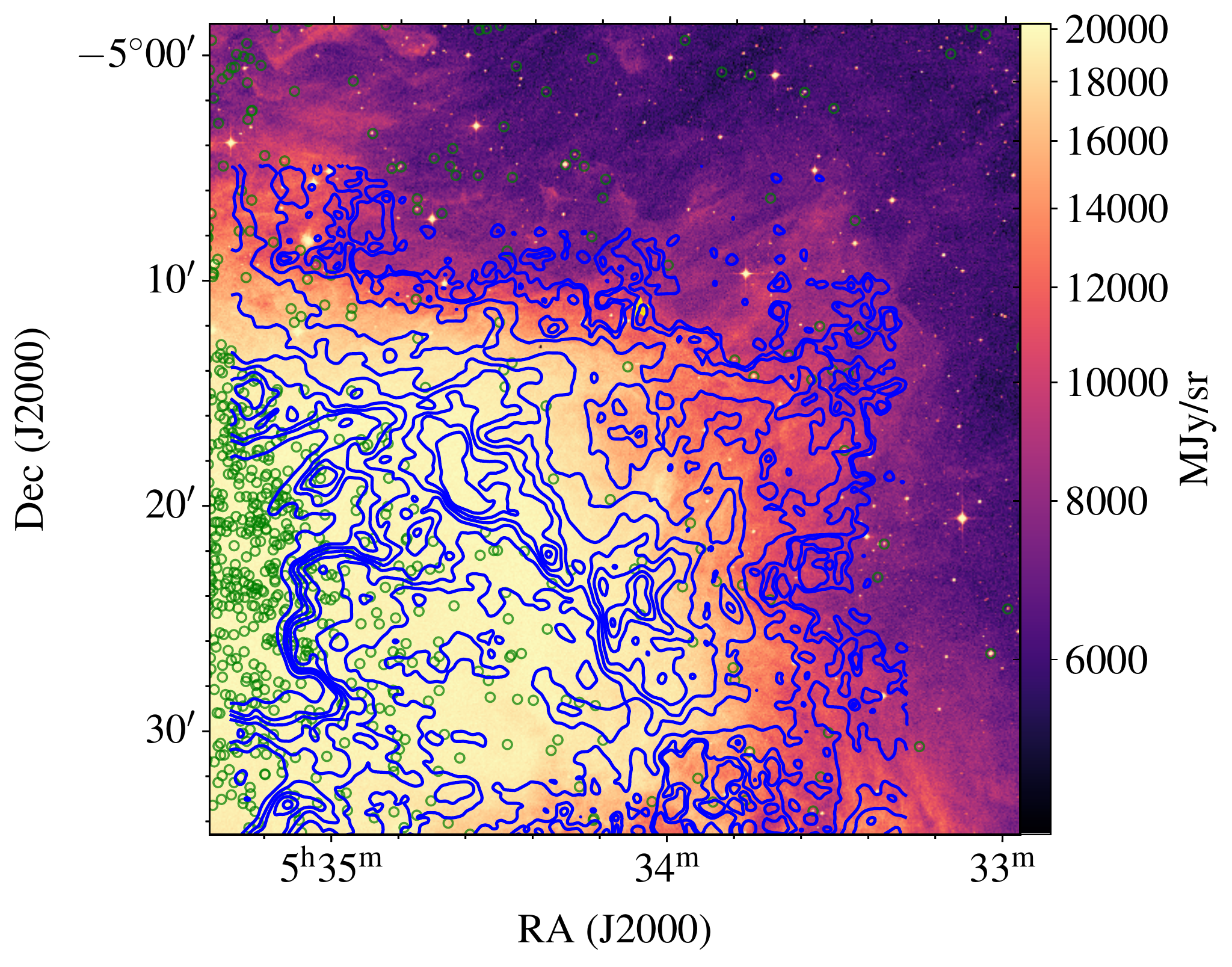}
        \caption{H${\alpha}$ image, which traces the ionized gas in the protrusion. The blue contours show the SOFIA \cii\,line integrated emission. Green circles show the young stars and protostars surveyed by \citet{Megeath2005, Megeath2012}.}
      \label{fig:m42_Halpha}
    \end{figure}

    Figs.~\ref{fig:m42_8micron} and \ref{fig:m42_Halpha} show \textit{Spitzer} 8~$\mu$m and H$\alpha$ maps, respectively. In both panels, blue contours are the integrated (between $-$5 and 14~km~s$^{-1}$) \cii\,observations. In Figs.~\ref{fig:m42_8micron}, \cii\,traces 8~$\mu$m closely. 
    
    \begin{figure*}[h!]
        \centering
        \includegraphics[width=0.55\hsize]{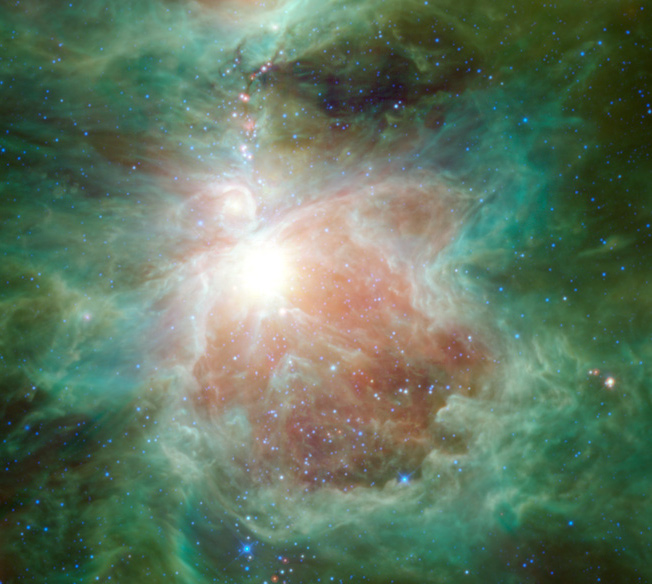}
        \caption{WISE image of the Orion Nebula provided by University of Berkeley. Blue represents emission at 3.4~$\mu$m and cyan (blue-green) represents 4.6~$\mu$m, both of which come mainly from hot stars. Relatively cooler objects, such as PAHs, the dust of the nebulae, appear green and red. Green represents 12~$\mu$m emission and red represents 22~$\mu$m emission tracing very small grains (VSGs). The field of view (FOV) of the original image is 3$^{\degr}$~$\times$~3$^{\degr}$, but we trimmed the image to show the outflow beyond the protrusion. The original file can be retrieved via: \url{http://wise.ssl.berkeley.edu/gallery_OrionNebula.html}}
        \label{fig:orion_wise_highCO}
    \end{figure*}
    
    \begin{figure*}[h!]
        \centering
        \includegraphics[width=0.4\hsize]{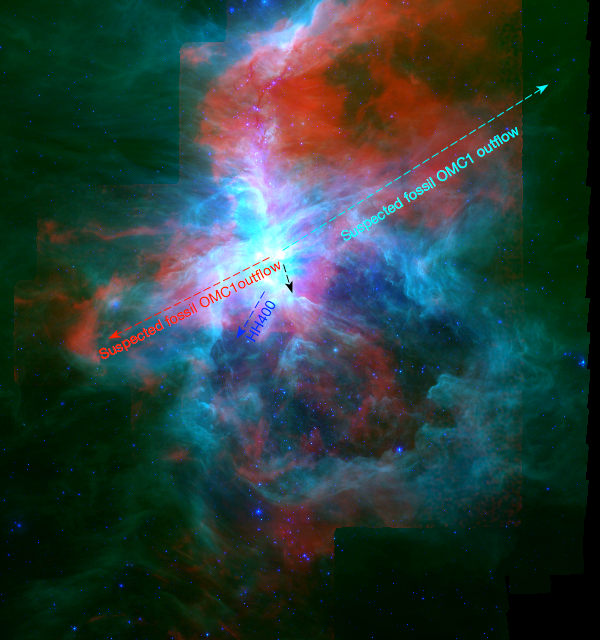}
        \includegraphics[width=0.55\hsize]{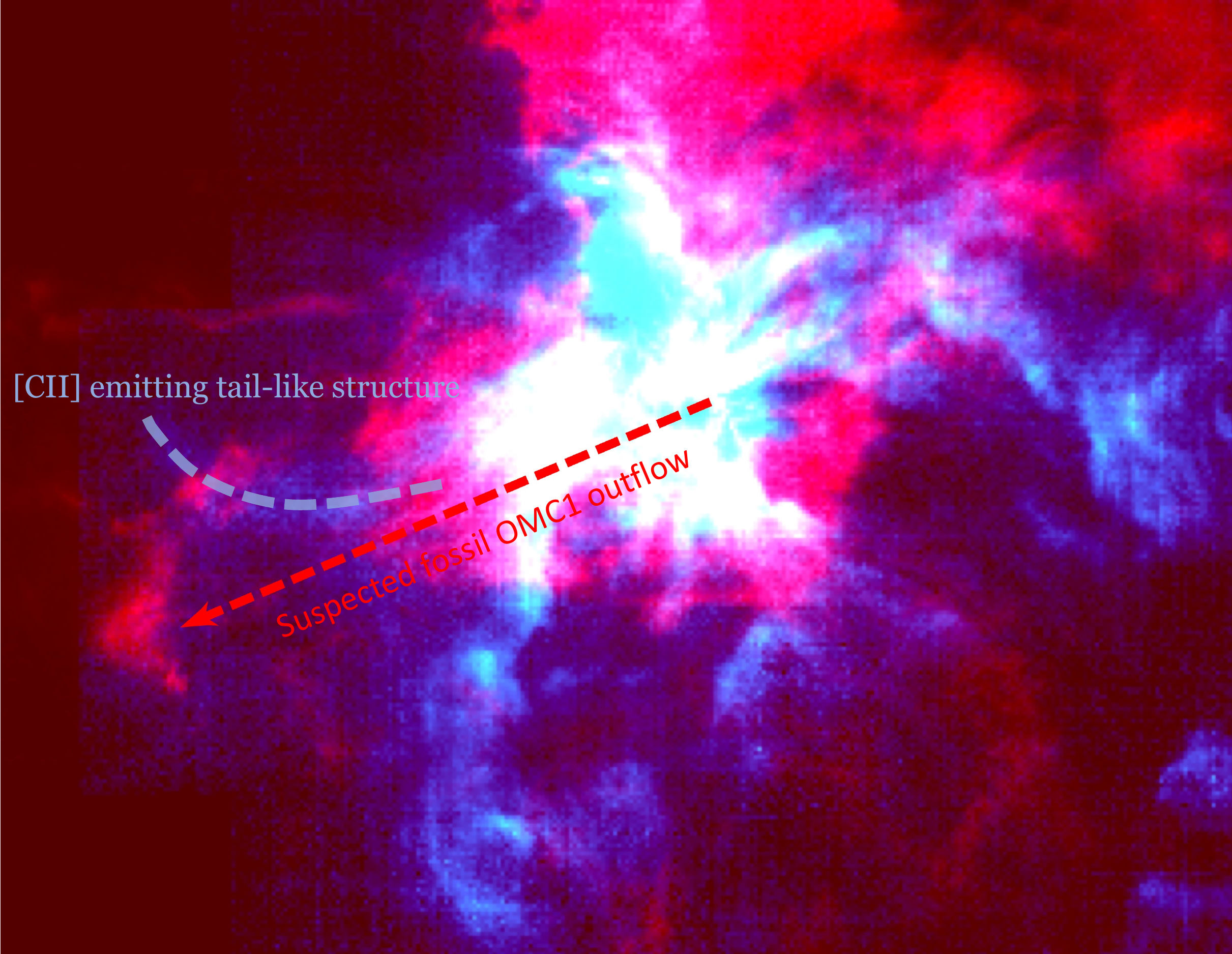}
        \caption{\textit{Left:} Red- and blue-shifted lobes of suspected fossil OMC-1 outflow on WISE image including CO-emission, which is red color (Bally et al., in prep.). \textit{Right:} Red-shifted lobe of suspected fossil OMC-1 outflow on \cii\, (blue emission) and $^{12}$CO emission (red emission). In both panels, CO emission is integrated between $+$10 and $+$13~km~s$^{-1}$.}
        \label{fig:orion_wise}
    \end{figure*}
    
    \begin{figure*}[h!]
        \centering
        \includegraphics[width=0.45\hsize]{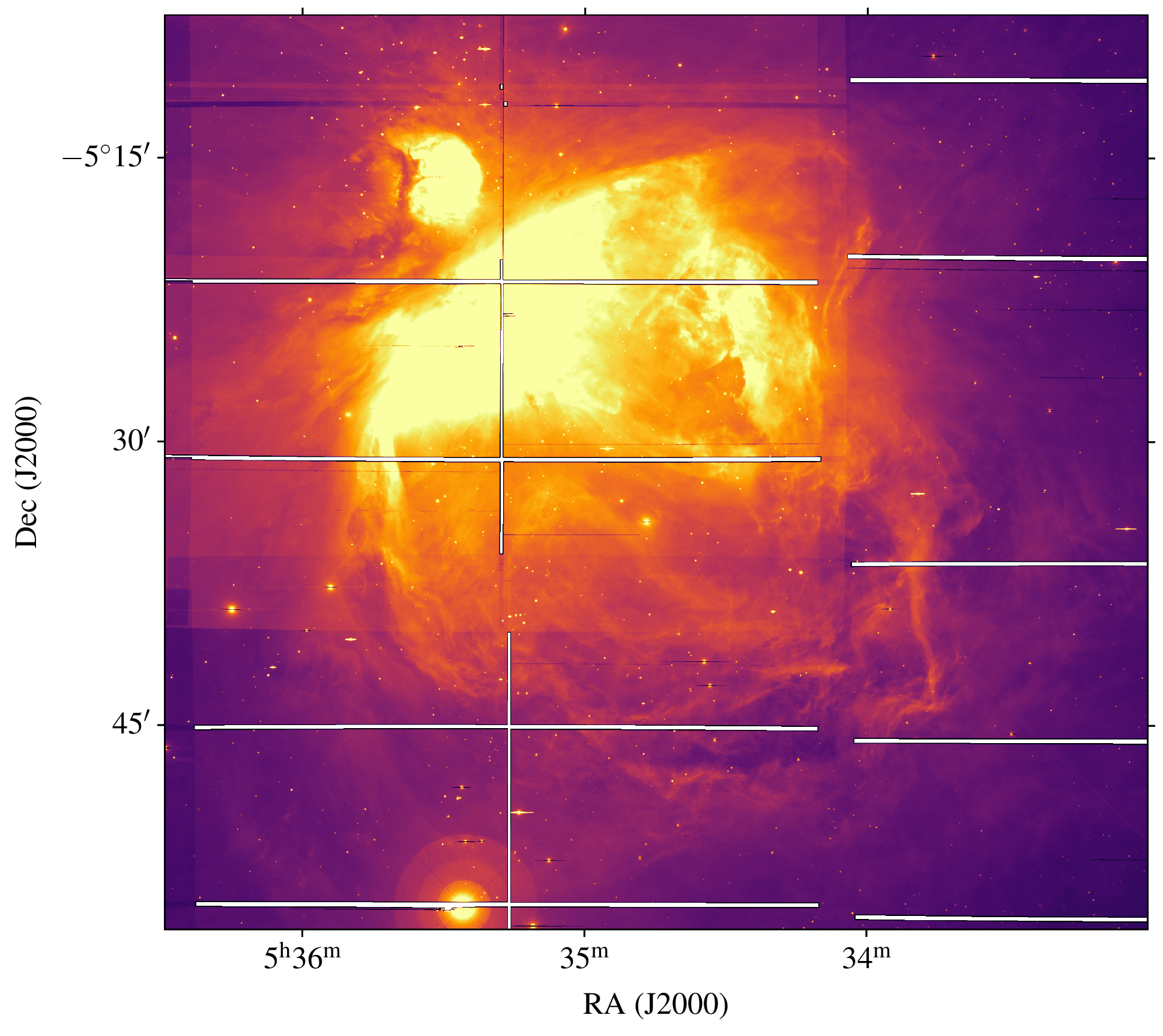}
        \caption{Orion \sii\,image, which has been taken with the V\'ictor M. Blanco 4-m Telescope in Chile, showing the giant outflow structures.}
        \label{fig:orion_sii}
    \end{figure*}
    
    \begin{figure*}[ht!]
       \centering
        \includegraphics[width=\hsize]{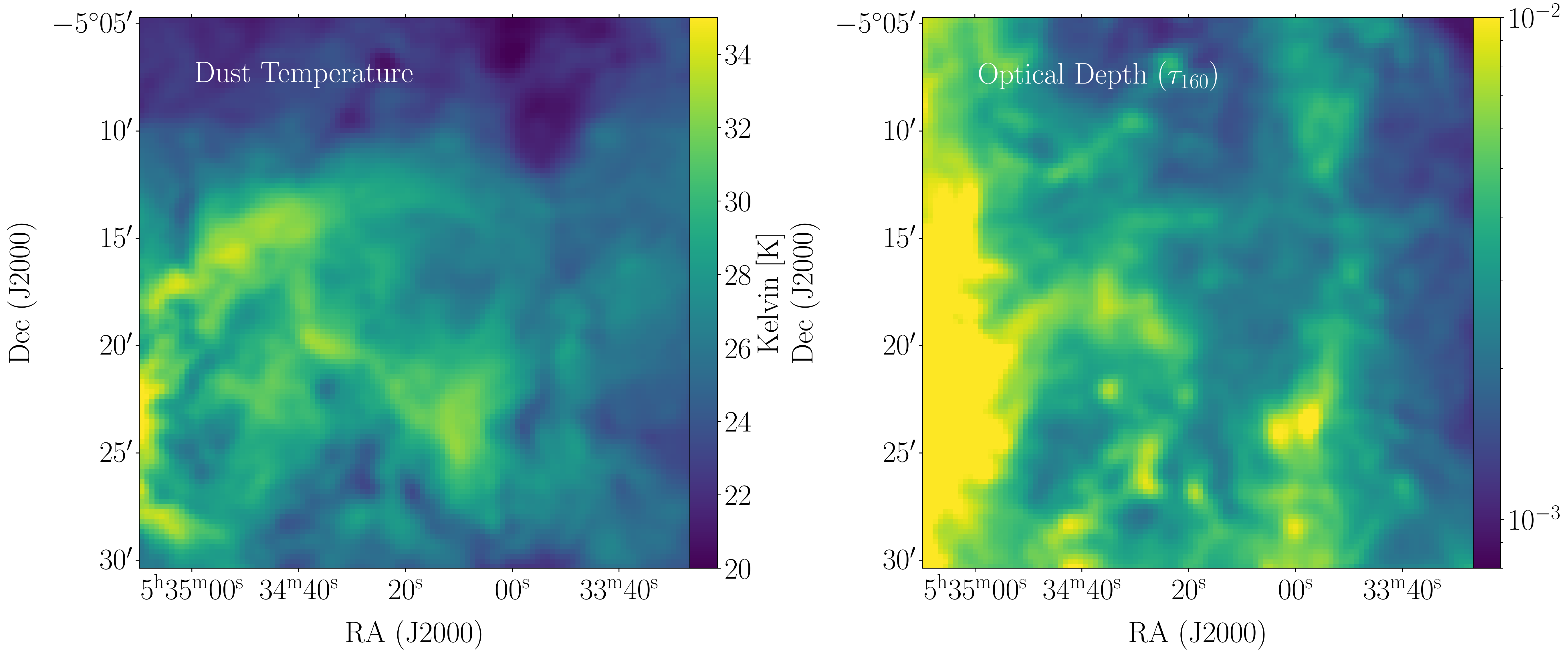}
        \includegraphics[width=0.5\hsize]{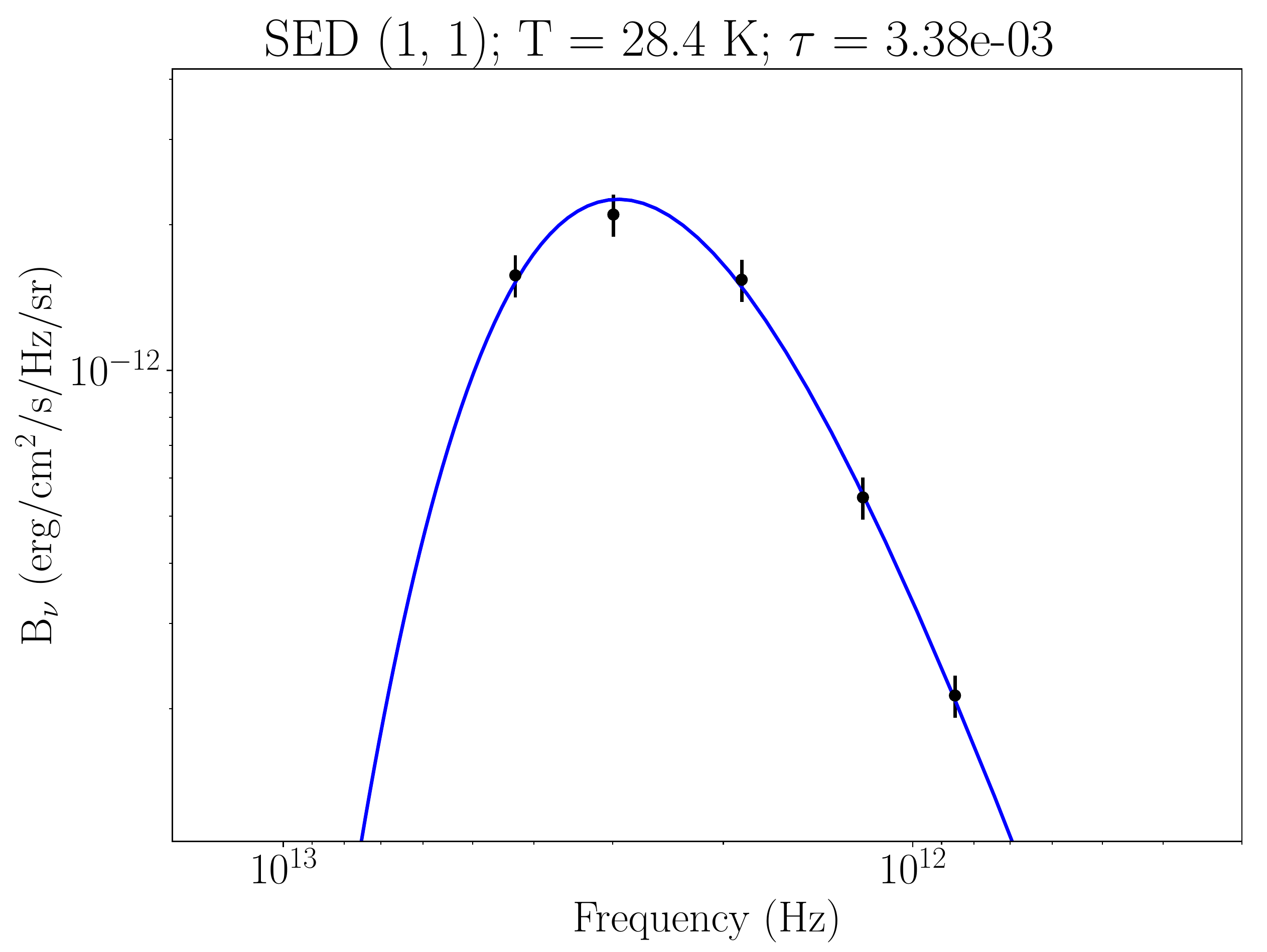}
        \caption{The temperature (upper-left) and the optical depth at 160~$\mu$m ($\tau_{160}$) (upper-right) map of dust emission, which traces the mass of the shell. The bottom panel shows an example of SEDs from the bottom-left of the dust temperature map.}
      \label{fig:dust_temp_depth}
    \end{figure*}

    \begin{figure*}[ht!]
       \centering
        \includegraphics[width=0.95\hsize]{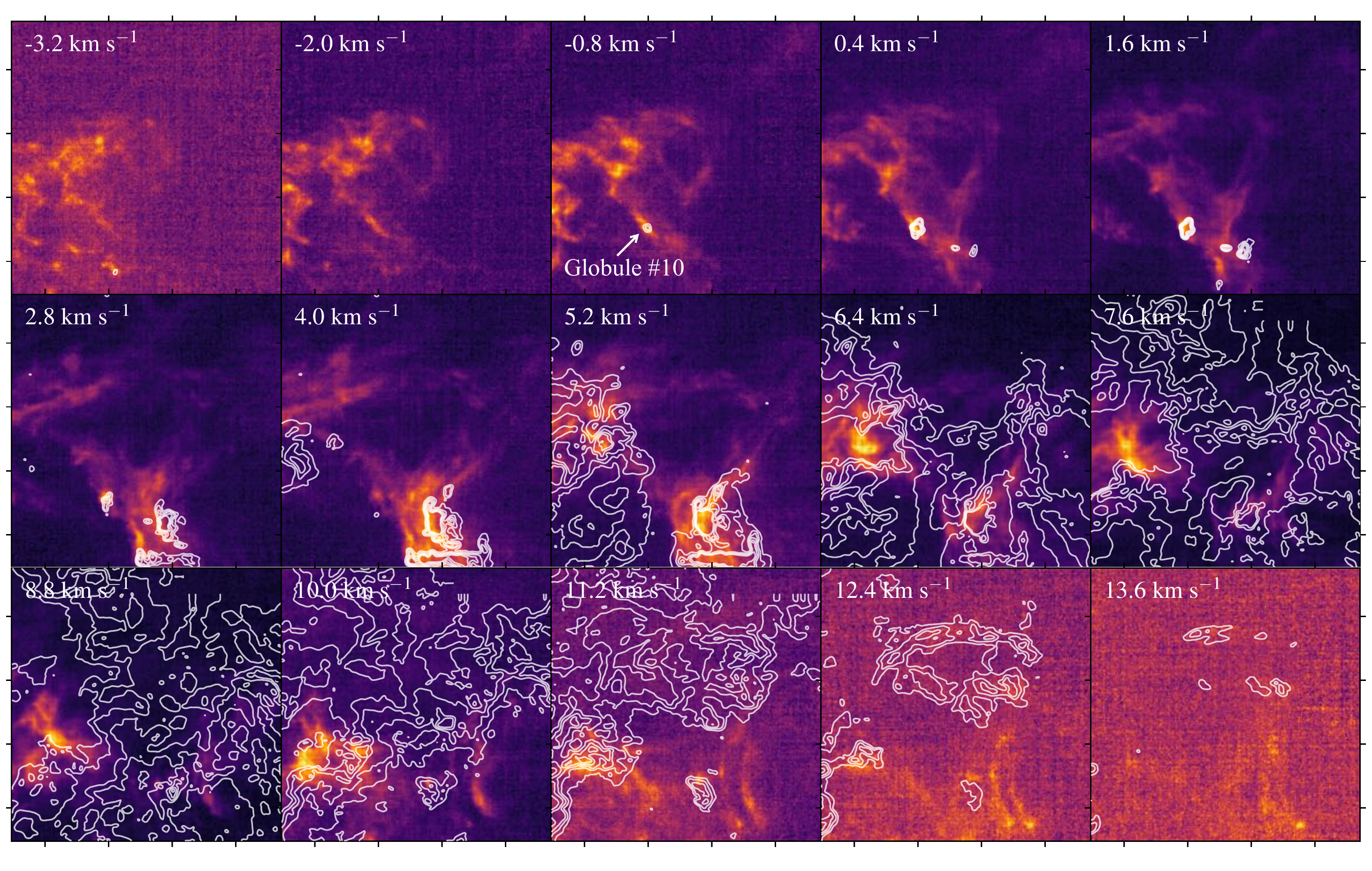}
        \caption{Channel map of \cii\,emission from $V_\mathrm{LSR}$ from $-$3.2 to $+$13.6~km~s$^{-1}$ overlaid with $^{12}$CO $J$ = 2-1 observations with white contours. The contour levels are [3, 6, 10, 15, 20]~K~km~s$^{-1}$. The velocity resolution of both maps is smoothed to 0.5~km~s$^{-1}$. Globule \#10 which is a bright CO emission at the velocity of $-$0.8~km~s$^{-1}$ indicates the CO globule reported in Orion Veil \citep[see also Fig.~\ref{fig:COPVdiagrams};][]{Goicoechea2020}.}
      \label{fig:PAH_cii_channelmaps3}
    \end{figure*}
    
    \begin{figure*}[ht!]
       \centering
        \includegraphics[width=\hsize]{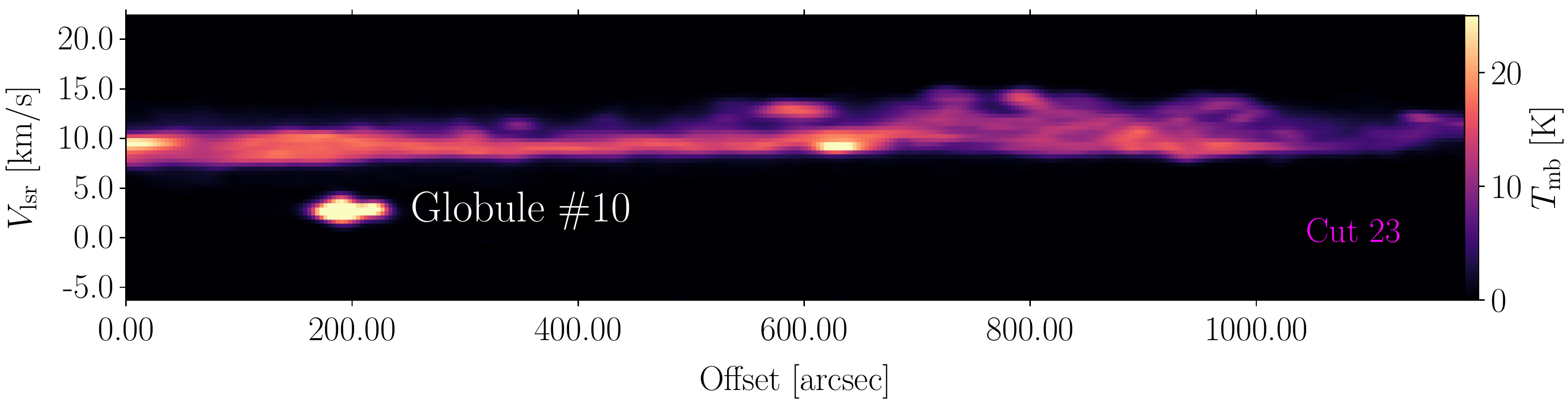}
        \caption{PV diagram of $^{12}$CO $J$ = 2-1 along the crosscut 23 in Fig.\ref{fig:pv_diagrams}. Globule \#10 which is a bright CO emission at 200$\arcsec$ indicates the CO globule reported by \citet{Goicoechea2020}.}
      \label{fig:COPVdiagrams}
    \end{figure*}
    
    \begin{figure}[ht!]
       \centering
        \includegraphics[width=.45\columnwidth]{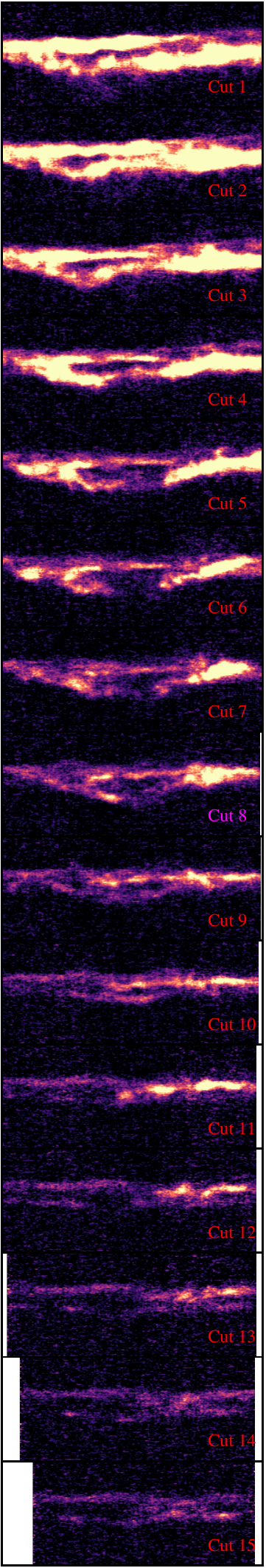}
        \caption{\cii\,pv diagrams from the protrusion sliced along expansion direction (i.e., cuts from 1 to 15 in Fig.~\ref{fig:pv_diagrams}). All diagram have same scales as in Fig.~\ref{fig:pv_diagrams}.}
      \label{fig:allPVdiagrams1}
    \end{figure}
    
    \begin{figure}[ht!]
       \centering
        \includegraphics[width=.45\columnwidth]{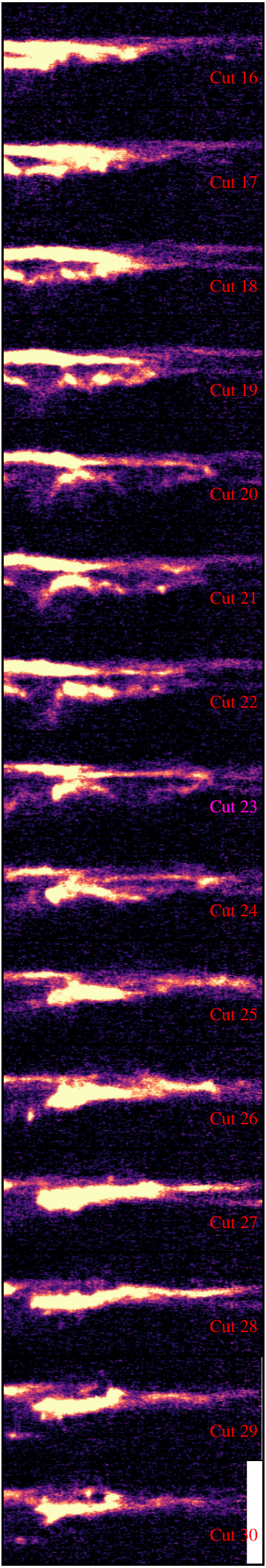}
        \caption{\cii\,pv diagrams from the protrusion sliced along expansion direction (i.e., cuts from 16 to 30 in Fig.~\ref{fig:pv_diagrams}). All diagram have same scales as in Fig.~\ref{fig:pv_diagrams}.}
      \label{fig:allPVdiagrams2}
    \end{figure}

    \begin{table*}[!ht]
        \centering
        \caption{List of O, B, and A stars within 0.5$\arcmin$ circle which is centered at Veil (RA: 83.6952553, Dec: -5.5075778) retrieved from SIMBAD. For object type, see \url{http://simbad.u-strasbg.fr/simbad/sim-display?data=otypes}. Star 39 is an A3 type star which has an luminosity of 14~$L_\sun$ and has mass from 1.4 to 2.1~$M_\sun$ on average. See Section~\ref{Sect:Results} for more detail.}
\begin{tabular}{lllllll}
\hline
{} &       ID &                 Main ID     &         RA         &       Dec            & Spectral Type &  Object Type \\
   &          &                             &       (J2000)      &      (J2000)         &               &              \\
\hline
1  &  811408   &    HD  37000               &  83.7958853993263  &  -5.9269107098082    &    B3/5       &        Y*O \\
2  &  813903   &    HD  37115               &  83.9753261999999  &  -5.6284203000000    &    B7Ve       &        Be* \\
3  &  812745   &    * iot Ori               &  83.8582579470833  &  -5.9099009825000    &    O9IIIvar   &        SB* \\
4  &  813805   &    HD  36960               &  83.7611697491666  &  -6.0020287608333    &    B1/2Ib/II  &          * \\
5  &  810070   &    * tet02 Ori B           &  83.8600017604396  &  -5.4168868162047    &    B2-B5      &        Y*O \\
6  &  800633   &    HD  37174               &  84.1132728798854  &  -5.4086975223920    &    B9V        &          * \\
7  &  810062   &    V* V1230 Ori            &  83.8363371224592  &  -5.3623168753733    &    B1         &        Or* \\
8  &  800723   &    * tet02 Ori C           &  83.8809635830516  &  -5.4212134690023    &    B4V        &        Or* \\
9  &  808906   &    Brun 328                &  83.6666476302733  &  -5.1686328311399    &    A0         &          * \\
10 &  800621   &    * tet01 Ori A           &  83.8159384038862  &  -5.3873145962644    &    B0V        &        Ae* \\
11 &  809750   &    2MASS J05355545-0513556 &  83.9810643811524  &  -5.2321419926502    &    A0-A5      &          * \\
12 &  800755   &    V* KO Ori               &  83.7353396045814  &  -5.5267106898729    &    A7         &        Or* \\
13 &  811404   &    HD  37150               &  84.0626137226411  &  -5.6479205223977    &    B3III/IV   &          * \\
14 &  800610   &    HD 294265               &  83.6436144295709  &  -5.0519180025009    &    A5         &          * \\
15 &  810606   &    V* V2254 Ori            &  83.8088040000000  &  -5.3729809999999    &    B          &        Or* \\
16 &  804756   &    HD  37061               &  83.8806889648567  &  -5.2673850427802    &    O9V        &        Or* \\
17 &  814861   &    * iot Ori B             &  83.8602322410440  &  -5.9123465817102    &    B8III      &        Or* \\
18 &  811401   &    HD  36999               &  83.8083635221351  &  -5.8267434770268    &    B8(III)    &        Y*O \\
19 &  800619   &    * tet01 Ori F           &  83.8196816666666  &  -5.3903246666666    &    B8         &        Em* \\
20 &  5385015  &    [AD95]  266             &  84.0075949582065  &  -5.5535633891067    &    A2-A7      &          * \\
21 &  803601   &    V* KS Ori               &  83.7505396759109  &  -5.4211711594684    &    A0V        &        Or* \\
22 &  811506   &    HD  36918               &  83.7046779606113  &  -6.0063687004004    &    B8.3       &          * \\
23 &  800625   &    V* MR Ori               &  83.8207427193823  &  -5.3625935531376    &    A2:Vv      &        Or* \\
24 &  805895   &    Brun 818                &  83.9174181358036  &  -5.2914985455163    &    B6         &          * \\
25 &  804750   &    * tet02 Ori A           &  83.8454260463393  &  -5.4160603284502    &    O9.5IVp    &        SB* \\
26 &  11673670 &    V* V566 Ori             &  83.8991213575002  &  -5.2057198015545    &    A0V        &        Or* \\
27 &  800617   &    * tet01 Ori D           &  83.8219059621791  &  -5.3879353377433    &    B1.5Vp     &        Y*O \\
28 &  801426   &    Brun 633                &  83.8297477353120  &  -5.3440962179628    &    A4-A7      &          * \\
29 &  810775   &    * tet01 Ori C           &  83.8185989772286  &  -5.3896801536110    &    O7Vp       &         ** \\
30 &  811405   &    HD  37188               &  84.1217809099165  &  -5.7701380790092    &    A7II/III   &         V* \\
31 &  804755   &    HD  36939               &  83.7304012218118  &  -5.5061373895440    &    B7/8II     &         V* \\
32 &  800613   &    BD-05  1309             &  83.7527946101197  &  -5.0857699544024    &    A0         &          * \\
33 &  810618   &    V* T Ori                &  83.9602015583414  &  -5.4763676043719    &    A3IVeb     &        Ae* \\
34 &  800325   &    HD  36917               &  83.6957640858897  &  -5.5707151968755    &    B9III/IV   &        Or* \\
35 &  811399   &    HD  36866               &  83.6396047588607  &  -5.7147765616562    &    A0III/IV   &        Y*O \\
36 &  813706   &    Brun 508                &  83.7651678114711  &  -5.9835702452170    &    B9V        &          * \\
37 &  811400   &    HD  36983               &  83.7817944568187  &  -5.8689996889292    &    B5(II/III) &        Y*O \\
38 &  801423   &    V* V2338 Ori            &  83.8283683591195  &  -5.2914159669635    &    A8-F0      &        Or* \\
39* &  803639   &    V* V2056 Ori           &  83.7082551983168  &  -5.3124025764650    &    A3         &        Or* \\
40 &  800750   &    HD  36982               &  83.7909855486891  &  -5.4647816038316    &    B1.5Vp     &        Or* \\
41 &  808571   &    HD  36981               &  83.7758266391769  &  -5.2044210543561    &    B7III/IV   &          * \\
42 &  806045   &    V* V1073 Ori            &  83.8684367155291  &  -5.4389859896652    &    B9.5V      &        Or* \\
43 &  800631   &    HD  37114               &  83.9939193101629  &  -5.3753809142824    &    B9V        &          * \\
44 &  800605   &    HD  36655               &  83.2811466381601  &  -5.3405835469492    &    B9V        &          * \\
45 &  802024   &    Brun 1018               &  84.1612500000000  &  -5.4727777777777    &    B6V        &          * \\
46 &  800623   &    HD  37019               &  83.8258625809957  &  -5.0651869253289    &    A0         &         V* \\
47 &  810053   &    * tet01 Ori B           &  83.8171333333333  &  -5.3852472222222    &    B1V        &        Al* \\
48 &  800614   &    BD-05  1310             &  83.7650833333333  &  -5.0956944444444    &    B9         &          * \\
%49 &  810115   &    NAME OrionBar D2        &  83.83916666666666 &  -5.4278055555555556 &    O6         &        HII \\
49 &  812778   &    BD-05  1300             &  83.6351315763823  &  -5.7627229481362    &    A3         &          * \\
50 &  800611   &    HD  36899               &  83.6761686688060  &  -5.1207163307963    &    A0V        &        Y*O \\
51 &  11682603 &    BD-05  1322             &  83.8590433332317  &  -5.8046204341559    &    A6Vn       &        Y*O \\
52 &  11680584 &    HD  36919               &  83.7028546483922  &  -5.9986011891444    &    B9V        &          * \\
53 &  811409   &    Brun 731                &  83.8716790464683  &  -5.9131225053704    &    A0         &      Y*O   \\
        \hline
        \end{tabular}
        \label{tab:OBA_lists}
    \end{table*}

    \end{appendix}

\end{document}